\documentclass{article}
\usepackage{slashed,verbatim,subfigure}
\usepackage[numbers,sort&compress]{natbib}
\usepackage{amsmath}
\usepackage{amssymb}
\usepackage{appendix}
\usepackage{graphicx}
\usepackage{dcolumn}
\usepackage{bm}
\usepackage[colorlinks=true,linktocpage=true,
linkcolor=blue,citecolor=blue]{hyperref}
\usepackage[a4paper]{geometry}
\newcommand{\be}{\begin{eqnarray}}
\newcommand{\ee}{\end{eqnarray}}
\begin{document}
\large
\title{\bf{Nonextensive effects on the viscous properties of hot and magnetized QCD matter}}
\author{Shubhalaxmi Rath\footnote{shubhalaxmi@iitb.ac.in}~~and~~Sadhana Dash\footnote{sadhana@phy.iitb.ac.in}\vspace{0.03in} \\ Department of Physics, Indian Institute of Technology Bombay, Mumbai 400076, India}
\date{}
\maketitle
\begin{abstract}
We have studied the effect of the nonextensive Tsallis mechanism on the viscous properties of hot QCD matter 
in the presence of a strong magnetic field. The results are compared to the case of absence of magnetic 
field. The viscous coefficients, such as the shear viscosity ($\eta$) and the bulk viscosity ($\zeta$) are determined in the similar environment by utilizing the nonextensive Tsallis mechanism within the relaxation time approximation of kinetic theory. We have observed that, when the nonextensive parameter $q$ is just above unity, both shear and bulk viscosities get increased as compared to their counterparts at $q=1$. This enhancement in viscosities is more evident in the additional presence of a strong magnetic field. Furthermore, some observables pertaining to the flow characteristic, fluid behavior and conformal symmetry of the medium are also explored. 

\end{abstract}

\newpage

\section{Introduction}
In recent years, there have been many phenomenological and theoretical developments on the understanding of the properties of extreme state of matter, {\em i.e.} quark-gluon plasma (QGP) which is produced in the initial stages of ultrarelativistic heavy ion collisions at Relativistic Heavy Ion Collider (RHIC) and Large Hadron Collider (LHC). Later, the evidence that the strong magnetic fields could be generated in noncentral events \cite{Skokov:IJMPA24'2009,Bzdak:PLB710'2012}, makes the study of such extreme state of matter relevant and interesting. It is well-known that in the noncentral heavy ion collision, two nuclei traveling with ultrarelativistic speed can be perceived as electric currents moving in opposite directions and when they collide, an extremely strong magnetic field perpendicular to the collision plane gets produced due to the relative motion between the colliding particles and the collision unaffected particles \cite{Shovkovy:LNP871'2013} and such magnetic field depends on the time, position, energies of the ions and impact parameter. The initially produced strong magnetic field is transient in nature, but, in the presence of finite electrical conductivity, the transient magnetic field induces a current which, in turn, elongates the lifetime of the strong magnetic field, consistent with the Lenz's law \cite{Tuchin:AHEP2013'2013,Rath:PRD100'2019}. So, it remains strong enough during the lifetime of QGP. The strength of the magnetic field rises approximately linearly with the center-of-mass energy ($\sqrt{s_{NN}}$) \cite{Bzdak:PLB710'2012,Tuchin:PRC88'2013} and its estimated magnitude ranges from $m_{\pi}^2$ ($\simeq 10^{18}$ Gauss) at RHIC to 15 $m_{\pi}^2$ at LHC \cite{Skokov:IJMPA24'2009,Bzdak:PLB710'2012}. Thus, the cumulative effect of high temperature, density and strong magnetic field could alter the properties of QGP and is being continued to be the topic of active research. The impact of magnetic field on various properties of hot and dense matter produced in heavy ion collisions has been studied in many works, such as the thermodynamic and magnetic properties \cite{Andersen:JHEP1210'2012,Strickland:PRD86'2012,Rath:JHEP1712'2017,
Bandyopadhyay:PRD100'2019,Rath:EPJA55'2019,Karmakar:PRD99'2019,Karmakar:PRD102'2020}, the transport properties \cite{Nam:PRD87'2013,Hattori:PRD94'2016,Feng:PRD96'2017,Hattori:PRD96'2017,Li:PRD97'2018,Denicol:PRD98'2018,
Fukushima:PRL120'2018,Rath:PRD100'2019,Das:PRD100'2019,Kurian:EPJC79'2019,Rath:PRD102'2020,Rath:EPJC80'2020,
Rath:EPJC81'2021,Rath:EPJC82'2022,Rath:EPJA59'2023}, the heavy quark diffusion \cite{Fukushima:PRD93'2016}, the dispersion relation in a magnetized thermal medium \cite{Sadooghi:PRD92'2015,Das:PRD97'2018}, the photon and dilepton productions from QGP \cite{Hees:PRC84'2011,Tuchin:PRC88'2013,Mamo:JHEP1308'2013,Shen:PRC89'2014,Bandyopadhyay:PRD94'2016,
Das:PRD99'2019} etc. The hot QCD matter is assumed to be a medium of quasiparticles with thermally generated masses. These masses may also be modulated by other energy scales arising due to the presence of the strong magnetic field and finite chemical potential in addition to the temperature scale, and thus affect the properties of hot QCD matter. Different statistical approaches, such as Fermi-Dirac, Bose-Einstein and Boltzmann-Gibbs approaches are frequently used in the study of properties of equilibrated systems. But, ideally the hot and dense matter produced in the aforesaid heavy ion collisions is not exactly in the locally equilibrated state and for a precise description of such matter, the nonextensive Tsallis approach can be treated as a relevant approach. This approach is well supported by the evidences of finding good fits of the transverse momentum spectra for a wide range of collision energies by STAR \cite{Abelev:PRC75'2007}, PHENIX \cite{Adare:PRC83'2011}, ALICE \cite{Aamodt:EPJC71'2011} and CMS \cite{Khachatryan:JHEP1105'2011} collaborations. To the best of our knowledge, no rigorous research on viscous properties, such as shear viscosity, bulk viscosity and associated observables has been carried out using the nonextensive Tsallis approach in the strong magnetic field regime. In the nonextensive Tsallis statistics, the parameter $q$ measures the extent of nonequilibration. Theoretical studies on the nonextensive statistics can be found in references \cite{Tsallis:JSP52'1988,Tsallis:BOOK'2009,Beck:EPJA40'2009,
Kaniadakis:EPJA40'2009,Kodama:EPJA40'2009,Alberico:EPJA40'2009,Biro:EPJA40'2009,
Alqahtani:EPJC82'2022}. In Langevin, Fokker-Planck, and Boltzmann type equations, $q=1$ represents the Boltzmann limit \cite{Walton:PRL84'2000,Kaniadakis:PLA288'2001,Biro:PRL94'2005,Sherman:LNP633'2004}. As the parameter $q$ deviates from the Boltzmann limit, temperature fluctuations were found to increase \cite{Cleymans:JPG36'2009}. 

The value of $q$ lies very close to 1. As per the fits of the RHIC and LHC spectra, $q$ for the hadronic matter varies in the range 1.08 - 1.2 \cite{Tang:PRC79'2009,Shao:JPG37'2010,Sikler:EPJWC13'2011} and 
for the quark matter, it deviates up to 1.22 \cite{Biro:JPG36'2009}. For $q>1$, the particle yields in the case of large $p_T$ can be the same for smaller value of the freeze-out temperature, whereas, the baryon chemical potential gets increased to compensate the decrease in the particle number with $q>1$ \cite{Cleymans:JPG36'2009}. As compared to the Boltzmann distribution, the chemical equilibrium was found to be more accurate in the case of the Tsallis distribution \cite{Cleymans:JPG36'2009}. The nonextensive effects are encoded in the dynamical model through the nonextensive Tsallis mechanism, and this mechanism is advantageous to understand the bulk properties of both QGP and hadronic mediums to a greater extent. The nonextensive parameter $q$ significantly affects the transverse momentum spectra, the multiplicity fluctuations, the nuclear modification factor etc. in heavy ion collisions \cite{Beck:PA322'2003,Biro:JPG35'2008,Tripathy:EPJA52'2016}. The deviation of parameter $q$ from unity could conspicuously influence various properties of the hot medium of quarks and gluons. In the present work, we study its effect on the viscous properties of the QCD medium. 

Slightly nonequilibrated system can possess finite shear ($\eta$) and bulk ($\beta$) viscosities. These viscosities hold essential characteristics in the hydrodynamic description of QGP. In general, 
viscosity is a measure of mutual friction between the surrounding fluid elements moving with distinct velocities. Basically, shear viscosity is associated with the momentum transfer in the medium and bulk viscosity explains the change of local pressure due to either expansion or contraction of fluid. In other words, the shear viscosity illustrates the resistance to any deformation in the system at constant volume and the bulk viscosity elucidates the resistance to change in the volume of the system at constant shape. The shear and bulk viscosities had been studied using different methods, such as the perturbation theory \cite{Arnold:JHEP11'2000,Arnold:JHEP05'2003,Arnold:PRD74'2006,Hidaka:PRD78'2008}, the kinetic theory \cite{Danielewicz:PRD31'1985,Sasaki:PRC79'2009,Heckmann:EPJA48'2012}, the lattice simulation \cite{Astrakhantsev:JHEP1704'2017,Astrakhantsev:PRD98'2018}, the molecular dynamics simulation \cite{Gelman:PRC74'2006}, the correlator technique using Green-Kubo formula \cite{Kharzeev:JHEP0809'2008,Moore:JHEP0809'2008,Plumari:PRC86'2012} etc. These viscosities could alter the properties of different observables in heavy ion collisions, such as the elliptic flow coefficient, hadron transverse momentum spectrum etc. \cite{Song:PLB658'2008,Denicol:JPG37'2010,Dusling:PRC85'2012,Noronha-Hostler:PRC90'2014}. The ratios of shear viscosity to entropy density ($\eta/s$) and bulk viscosity to entropy density ($\zeta/s$) are also very important quantities in estimating the phase transition point of matter \cite{Csernai:PRL97'2006}. Ratio $\eta/s$ delineates the fluid behavior of matter, whereas ratio $\zeta/s$ is useful in discerning the conformal symmetry of the system. According to the estimations on the results for Yang-Mills theory and perturbative QCD \cite{Arnold:JHEP05'2003}, there is a slight increase of the ratio $\eta/s$ in the presence of dynamical quarks \cite{Astrakhantsev:JHEP1704'2017,Meyer:NPA830'2009}. Alternatively, the shear viscosity was determined by exploiting the functional diagrammatic approaches to QCD through the Kubo relation from gluon spectral functions \cite{Haas:PRD90'2014,Christiansen:PRL115'2015}. The results reported in ref. \cite{Christiansen:PRL115'2015} are closer to the lattice results. Both the ﬁrst-principle approaches ﬁnd the value of $\eta/s$ of about 0.2 near $T_c$ with an increase with $T$ and this is in good agreement with the estimations of ﬂuid dynamical simulations. For a strongly interacting matter with finite shear and bulk viscosities, the experimental data on the multiplicity, the transverse momentum spectra and the elliptic flow are accurately explained in ref. \cite{Ryu:PRL115'2015}. As per the anti-de sitter/conformal field theory (AdS/CFT), lower bound of $\eta/s$ is $1/(4\pi)$, where matter behaves like a strongly interacting perfect fluid \cite{Kovtun:PRL94'2005}. Perturbative QCD calculations have predicted comparatively higher value of $\eta/s$ \cite{Arnold:PRD74'2006}. For a system with conformal symmetry, $\zeta/s$ approaches zero. Further, the lattice simulations \cite{Borsonyi:JHEP11'2010,Bazavov:PRD90'2014} suggest that the QCD does not show conformal symmetry in the vicinity of the critical temperature ($T_c$) due to the appearance of a peak in the trace anomaly $(\epsilon-3P)/T^4$, where $\epsilon$ and $P$ are respectively the energy density and the pressure of the medium. 

The presence of strong magnetic field modifies the dispersion relation and for $f$th flavor of quark with absolute charge $|q_f|$ and mass $m_f$, this relation is written as $\omega_{f,n}=\sqrt{p_L^2+2n|q_fB|+m_f^2}$, where $p_L$ represents the longitudinal component of momentum with respect to the direction of magnetic field and the transverse component of the momentum ($p_T$) is quantized in terms of the Landau levels ($n$). In the strong magnetic field regime, magnetic field scale is much larger than the temperature scale, {\em i.e.} $|q_fB| \gg T^2$. In this regime, the energy gap between Landau levels is very large ($\sim\mathcal{O}(\sqrt{|q_fB|})$), thus the quarks could not jump to the higher Landau levels and occupy only the lowest Landau levels (LLL). However, the gluons being electrically neutral particles do not get directly affected by the strong magnetic field, but can get indirectly affected through their thermal masses. In the presence of magnetic field, the rotational symmetry gets broken and this makes the viscous stress tensor to split into five shear viscous coefficients and two bulk viscous coefficients \cite{Lifshitz:BOOK'1981,Tuchin:JPG39'2012,Critelli:PRD90'2014,Hernandez:JHEP05'2017,Hattori:PRD96'2017,
Chen:PRD101'2020}. Specifically in the strong magnetic field limit, the components of shear and bulk viscosities along the direction of magnetic field only exist \cite{Lifshitz:BOOK'1981,Tuchin:JPG39'2012,Hattori:PRD96'2017}. Now, it would be interesting to see how the nonextensive behavior of hot QCD medium alters the values of $\eta$, $\zeta$, $\eta/s$ and $\zeta/s$. After knowing that, it would be possible to trace the information on how far the nonextensive medium appears from an ideal hydrodynamics. Our recent observation \cite{Rath:EPJC83'2023} showed that the effect of the nonextensivity on different charge and heat conductivities is predominant in the presence of finite magnetic field. So, the viscous properties might also be greatly influenced by the nonextensive behavior of the medium and this nonextensive effect could be more illuminating in the presence of a strong magnetic field. The present work intends to study the viscous properties of the hot QCD medium using the quasiparticle model for strong magnetic field and finite chemical potential. We further intend to study the collective effects due to nonextensivity and the strong magnetic field on some observables pertaining to the flow characteristic, fluid behavior and conformal symmetry of the QCD medium. 

In this work, we have used the relativistic Boltzmann transport equation in a nonextensive hot QCD medium within the kinetic theory approach. In this transport equation, the nonextensive behavior is encoded 
in the Tsallis form of the particle distribution function containing $q$ parameter. For the calculation of the transport coefficients, we have solved the relativistic Boltzmann transport equation by using the relaxation time approximation. In some previous works, Tsallis mechanism had been applied to both Fokker-Planck equation and Plastino-Plastino equation, and comparative studies had been presented. In general, the Fokker-Planck equation is a second order approximation of the Boltzmann equation, and it can be used as a basic tool for the exploration of kinematic aspects of a variety of systems. It avoids off the integral equation by the introduction of transport coefficients, such as the drift term and the diffusion term. On the other hand, the Plastino-Plastino equation is a generalization describing the kinematic evolution of complex systems consistent with the $q$-statistics. For example, in systems governed by the non-additive entropy, the Plastino-Plastino equation was used as a generalization of the Fokker-Planck equation in ref. \cite{Plastino:PA222'1995}, which gets reduced to the Fokker-Planck equation when the entropic index $q=1$. Later, the Plastino-Plastino equation has been further explored in many works \cite{Lutz:NP9'2013,Combe:PRL115'2015,Gomez:PRE107'2023}. A comparative study of the heavy-quark dynamics with the Fokker-Planck equation and the Plastino-Plastino equation has been demonstrated recently in ref. \cite{Eugenio:PLB845'2023}. In ref. \cite{Deppman:PLB839'2023}, the authors have studied the formal connections between the nonlinear Fokker-Planck equation associated with the non-additive entropy and the Boltzmann equation with the non-additive correlation functional, where the collision term following the $q$-algebra has been adopted, which is different from the relaxation time approximation used in the collision term in our work. 

Recently, some investigations have uncovered a strong correlation between the nonextensive statistics and fractal structures. It has been observed that the dynamics of systems within fractal spaces lead to $q$-exponential distributions, where the value of the entropic index $q$ is determined by the parameters of the fractal geometry. In ref. \cite{Klepikov:MPLB34'2020}, the influence of the fractality of a medium on the dynamics of charged particles in the external magnetic field has been considered and they reported a significant increase in the sensitivity to weak effects with the fractality of a medium. They have also analyzed the dynamical processes in the nonequilibrium nonconservative medium in the framework of the Tsallis nonextensive thermodynamics. The study in ref. \cite{Wang:PA340'2004} has shown the derivation of thermodynamics from Tsallis entropy for nonadditive systems. Further, the work in ref. \cite{Deppman:PO16'2021} has reported that the accurate description of the information dynamics in a fractal network is given by the $q$-exponential function in the Tsallis statistics. In references \cite{Deppman:AHEP2018'2018,Deppman:PRD101'2020}, the link between fractals, nonextensive Tsallis statistics and renormalization group invariance of Yang-Mills theory has been described. The results were applied to calculate $q$ for QCD in the one-loop approximation, which showed a good agreement with the value obtained experimentally. 

There also exists a wide range of applications of the nonextensive Tsallis statistics in quantum chromodynamics. A review on the applications of fractal structures of Yang-Mills fields 
and the nonextensive Tsallis statistics in high energy physics, including physics at Large Hadron Collider, hadron physics and neutron stars can be found in ref. \cite{Deppman:MDPIP2'2020}. Recently, the applications of the nonextensive Tsallis statistics to QCD and high energy physics have been studied in references \cite{Deppman:AHEP2018'2018,Deppman:PRD101'2020} and they reported possible connections of this statistics with the fractal structure of hadrons. The nonextensive parameter $q$ was deduced in terms of the field theory parameters, and this resulted in a good agreement with the experimental observations. The application of the nonextensive self-consistent thermodynamics has extended to systems with finite chemical potential and this helps to study the thermodynamic properties of the hadrons, the neutron stars etc. \cite{Lavagno:EPJA47'2011,Menezes:EPJA51'2015}. It was observed that the internal temperature of the neutron stars decreases with the increase of the nonextensive parameter. The applications of the nonextensive Tsallis distributions in high energy physics \cite{Cleymans:JPG39'2012,Sena:EPJA49'2013,Wilk:PLB727'2013,Wong:PRD91'2015}, hadron physics \cite{Deppman:JPG41'2014,Menezes:EPJA51'2015,Deppman:JPCS607'2015} etc. have motivated the formulation of thermofractals. The thermofractal theory is useful in discerning the hadron structure and some models have used the Tsallis statistics to introduce the fractal aspects of QCD in the description of hadron structure \cite{Cardoso:EPJA53'2017,Andrade:PRD101'2020}. 

The present work is organized as follows. Section 2 focuses on the effects of the nonextensivity on the shear and bulk viscosities for a hot QCD medium in the absence as well as in the presence of a strong magnetic field. The results are discussed in section 3. Some applications of the aforesaid viscous properties, such as the flow characteristic, the specific shear viscosity and the specific bulk viscosity are studied in section 4. Finally, in section 5, the results are concluded. 

\section{Shear and bulk viscous properties}
In this section, we are going to study the shear and bulk viscous properties of hot QCD medium using the relaxation time approximation within the Tsallis nonextensive framework in the absence of magnetic 
field and in the presence of a strong magnetic field in subsections 2.1 and 2.2, respectively. 

\subsection{Nonextensive hot QCD medium in the absence of magnetic field}
To study shear viscosity, bulk viscosity and other related phenomena under the influence 
of the nonextensivity, we adopt a nonextensive QCD medium with Tsallis formalism \cite{Conroy:PRD78'2008,Biro:PRC85'2012,Cleymans:JPG39'2012}. In this formalism, the fermion distribution function is written as
\be\label{Q.D.F.}
f=\frac{1}{\left[1+(q-1)\beta(u^\alpha p_\alpha\mp\mu_f)\right]^{\frac{1}{q-1}}+1}
~,\ee
where `$-$' sign is for quark case ($f_q$), `$+$' sign is for antiquark case ($\bar{f}_q$), $q$ represents the nonextensive parameter, $p_\alpha\equiv\left(\omega_f,\mathbf{p}\right)$, $\omega_f=\sqrt{\mathbf{p}^2+m_f^2}$, $u^\alpha$ denotes the 
four-velocity of fluid, $T=\beta^{-1}$ and $\mu_f$ is the chemical potential of the $f$th 
flavor of quark. In the aforesaid framework, the gluon distribution function is represented as
\be\label{G.D.F.}
f_g=\frac{1}{\left[1+(q-1)\beta u^\alpha p_\alpha\right]^{\frac{1}{q-1}}-1}
~,\ee
where $p_\alpha\equiv\left(\omega_g,\mathbf{p}\right)$. The deviation of $q$ from unity 
explains the extent of the nonextensivity of the system, {\em i.e.} the deviation 
of the medium from the equilibrated thermal distribution of particles. As $q$ approaches 1, the abovementioned nonextensive distribution functions can be approximated to Fermi-Dirac (for fermions) and Bose-Einstein (for bosons) distribution functions. However, the nonextensive systems can also get further deviated from equilibrium due to the action of external forces or fields, for example, an infinitesimal shift of the nonextensive distribution function from its near-equilibrium state can be considered under such circumstance. Thus, due to the action of external force, there is an infinitesimal shift in the energy-momentum tensor as well as in the particle distribution function, so, $T^{\mu\nu}$ changes to ${T^\prime}^{\mu\nu}=T^{\mu\nu}+\Delta T^{\mu\nu}$ and also, $f_q$, $\bar{f}_q$ and $f_g$ respectively change to $f_q^\prime=f_q+\delta f_q$, $\bar{f}_q^\prime=\bar{f}_q+\delta \bar{f}_q$ and $f_g^\prime=f_g+\delta f_g$ with $\delta f_q$, $\delta \bar{f}_q$ and $\delta f_g$ respectively representing the infinitesimal changes in quark, antiquark and gluon distribution functions. The energy-momentum tensor for a slightly nonequilibrium system is defined as
\be
{T^\prime}^{\mu\nu}=\int\frac{d^3{\rm p}}{(2\pi)^3}p^\mu p^\nu \left[\sum_f g_f\frac{\left(f_q^\prime+\bar{f}_q^\prime\right)}{{\omega_f}}+g_g\frac{f_g^\prime}{\omega_g}\right]
,\ee
where $g_f$ and $g_g$ are the degeneracy factors for quark and gluon, respectively and the flavor index $f$ takes the flavors $u$, $d$ and $s$. Similarly, the infinitesimal shift of the energy-momentum tensor in a nonequilibrium medium is given by
\be\label{em1}
\Delta T^{\mu\nu}=\int\frac{d^3{\rm p}}{(2\pi)^3}p^\mu p^\nu \left[\sum_f g_f\frac{\left(\delta f_q+\delta \bar{f}_q\right)}{{\omega_f}}+g_g\frac{\delta f_g}{\omega_g}\right]
.\ee

The infinitesimal change of the particle distribution function can be obtained by solving the relativistic Boltzmann transport equation in the relaxation time approximation. This approximation 
can also be applied to systems, which are infinitesimally deviated from equilibrium. We note that the nonextensive systems can also achieve equilibrium and some quasi-stationary state systems near equilibrium can be accurately described using Tsallis statistics, even if the equilibrium is described by Fermi-Dirac statistics (for fermions) and Bose-Einstein statistics (for bosons). This had been witnessed earlier in some works related to the nonextensive systems \cite{Song:PLB658'2008,Biro:PRC85'2012,Tripathy:EPJA52'2016,Alqahtani:EPJC82'2022}. In general, the Boltzmann transport equation is a complicated nonlinear integro-differential equation for particle distribution function, which gets linearized through the relaxation time approximation. This approximation is one of the frequently used methods to simplify the collision term through the following assumptions: (i) The distribution function gets infinitesimally deviated from its equilibrium, so that within a phenomenological timescale (the relaxation time) $\tau$, the system returns back to the equilibrium state. (ii) The probability per unit time for a collision, {\em i.e.}, ${1}/{\tau}$ does not depend on the parton distribution function. (iii) The number of partons scattering into the phase space volume element, involves the equilibrium distribution function, whereas the number of partons moving out of the concerned phase space volume element after suffering collisions, involves the nonequilibrium distribution function. Thus, in the relaxation time approximation, the collision term retains its simple form even for the nonextensive distribution functions. Thus, in order to calculate the infinitesimal changes of the nonextensive Tsallis distribution functions for quark, antiquark and gluon, we use their respective relativistic Boltzmann transport equations in the relaxation time approximation as
$p^\mu\partial_\mu f_q^\prime(x,p)=-\frac{p_\nu u^\nu}{\tau_f}\delta f_q(x,p),$ 
$p^\mu\partial_\mu \bar{f}_q^\prime(x,p)=-\frac{p_\nu u^\nu}{\tau_{\bar{f}}}\delta \bar{f}_q(x,p),$  $p^\mu\partial_\mu f_g^\prime(x,p) = -\frac{p_\nu u^\nu}{\tau_g}\delta f_g(x,p).$ 
The relaxation times for quarks (antiquarks), $\tau_f$ ($\tau_{\bar{f}}$) and gluons, $\tau_g$ are written \cite{Hosoya:NPB250'1985} as
\begin{eqnarray}
&&\tau_{f(\bar{f})}=\frac{1}{5.1T\alpha_s^2\log\left(1/\alpha_s\right)\left[1+0.12 
(2N_f+1)\right]} ~, \label{Q.R.T.} \\
&&\tau_g=\frac{1}{22.5T\alpha_s^2\log\left(1/\alpha_s\right)\left[1+0.06N_f
\right]} \label{G.R.T.}
~,\end{eqnarray}
respectively. Now, eq. \eqref{em1} turns out to be
\be\label{em2}
\Delta T^{\mu\nu}=-\int\frac{d^3{\rm p}}{(2\pi)^3}\frac{p^\mu p^\nu}{p_\nu u^\nu} \left[\sum_f g_f\frac{\left(\tau_f p^\mu\partial_\mu f_q^\prime+\tau_{\bar{f}} p^\mu\partial_\mu \bar{f}_q^\prime\right)}{\omega_f}+g_g\frac{\tau_g p^\mu\partial_\mu f_g^\prime}{\omega_g}\right]
,\ee
where the partial derivative is given by $\partial_\mu=u_\mu D+\nabla_\mu$, with $D=u^\mu\partial_\mu$. In the local rest frame, one can expand the distribution functions 
in terms of the gradients of flow velocity and temperature. The partial 
derivatives of the nonextensive quark, antiquark and gluon distribution 
functions are respectively calculated as
\begin{eqnarray}
\nonumber\partial_\mu f_q^\prime = \frac{\beta\left[1+(q-1)\beta(u_\alpha p^\alpha-\mu_f)\right]^{\frac{2-q}{q-1}}}{\left(\left[1+(q-1)\beta(u_\alpha p^\alpha-\mu_f)\right]^\frac{1}{q-1}+1\right)^2}\left[u_\alpha p^\alpha u_\mu\frac{DT}{T}+u_\alpha p^\alpha\frac{\nabla_\mu T}{T}-u_\mu p^\alpha Du_\alpha\right. \\ \left.-p^\alpha\nabla_\mu u_\alpha+T\partial_\mu\left(\frac{\mu_f}{T}\right)\right]
,\end{eqnarray}
\begin{eqnarray}
\nonumber\partial_\mu \bar{f}_q^\prime = \frac{\beta\left[1+(q-1)\beta(u_\alpha p^\alpha+\mu_f)\right]^{\frac{2-q}{q-1}}}{\left(\left[1+(q-1)\beta(u_\alpha p^\alpha+\mu_f)\right]^\frac{1}{q-1}+1\right)^2}\left[u_\alpha p^\alpha u_\mu\frac{DT}{T}+u_\alpha p^\alpha\frac{\nabla_\mu T}{T}-u_\mu p^\alpha Du_\alpha\right. \\ \left.-p^\alpha\nabla_\mu u_\alpha-T\partial_\mu\left(\frac{\mu_f}{T}\right)\right]
,\end{eqnarray}
\begin{eqnarray}
\partial_\mu f_g^\prime = \frac{\beta\left[1+(q-1)\beta u_\alpha p^\alpha\right]^{\frac{2-q}{q-1}}}{\left(\left[1+(q-1)\beta u_\alpha p^\alpha\right]^\frac{1}{q-1}-1\right)^2}\left[u_\alpha p^\alpha u_\mu\frac{DT}{T}+u_\alpha p^\alpha\frac{\nabla_\mu T}{T}-u_\mu p^\alpha Du_\alpha-p^\alpha\nabla_\mu u_\alpha\right]
.\end{eqnarray}
Substituting the expressions of $\partial_\mu f_q^\prime$, $\partial_\mu \bar{f}_q^\prime$ and $\partial_\mu f_g^\prime$ in eq. \eqref{em2} and then utilizing $\frac{DT}{T}=-\left(\frac{\partial P}{\partial \varepsilon}\right)\nabla_\alpha u^\alpha$ and $Du_\alpha=\frac{\nabla_\alpha P}{\varepsilon+P}$ from the energy-momentum conservation, we have
\be\label{em3}
\nonumber\Delta T^{\mu\nu} &=& \sum_f g_f\int\frac{d^3{\rm p}}{(2\pi)^3}\frac{p^\mu p^\nu}{\omega_f T}\left[\frac{\tau_f\left[1+(q-1)\beta(u_\alpha p^\alpha-\mu_f)\right]^{\frac{2-q}{q-1}}}{\left(\left[1+(q-1)\beta(u_\alpha p^\alpha-\mu_f)\right]^\frac{1}{q-1}+1\right)^2}\left\lbrace\omega_f\left(\frac{\partial P}{\partial \varepsilon}\right)\nabla_\alpha u^\alpha\right.\right. \\ && \left.\left.\nonumber+p^\alpha\left(\frac{\nabla_\alpha P}{\varepsilon+P}-\frac{\nabla_\alpha T}{T}\right)-\frac{Tp^\alpha}{\omega_f}\partial_\alpha\left(\frac{\mu_f}{T}\right)+\frac{p^\alpha p^\beta}{\omega_f}\nabla_\alpha u_\beta\right\rbrace\right. \\ && \left.\nonumber+\frac{\tau_{\bar{f}}\left[1+(q-1)\beta(u_\alpha p^\alpha+\mu_f)\right]^{\frac{2-q}{q-1}}}{\left(\left[1+(q-1)\beta(u_\alpha p^\alpha+\mu_f)\right]^\frac{1}{q-1}+1\right)^2}\left\lbrace\omega_f\left(\frac{\partial P}{\partial \varepsilon}\right)\nabla_\alpha u^\alpha\right.\right. \\ && \left.\left.\nonumber+p^\alpha\left(\frac{\nabla_\alpha P}{\varepsilon+P}-\frac{\nabla_\alpha T}{T}\right)+\frac{Tp^\alpha}{\omega_f}\partial_\alpha\left(\frac{\mu_f}{T}\right)+\frac{p^\alpha p^\beta}{\omega_f}\nabla_\alpha u_\beta\right\rbrace\right] \\ && \nonumber+g_g\int\frac{d^3{\rm p}}{(2\pi)^3}\frac{p^\mu p^\nu}{\omega_g T}\frac{\tau_g\left[1+(q-1)\beta u_\alpha p^\alpha\right]^{\frac{2-q}{q-1}}}{\left(\left[1+(q-1)\beta u_\alpha p^\alpha\right]^\frac{1}{q-1}-1\right)^2}\left[\omega_g\left(\frac{\partial P}{\partial \varepsilon}\right)\nabla_\alpha u^\alpha\right. \\ && \left.+p^\alpha\left(\frac{\nabla_\alpha P}{\varepsilon+P}-\frac{\nabla_\alpha T}{T}\right)+\frac{p^\alpha p^\beta}{\omega_g}\nabla_\alpha u_\beta\right]
.\ee
One can obtain the pressure and the energy density from the energy-momentum tensor as $P=-\Delta_{\mu\nu}T^{\mu\nu}/3$ and $\varepsilon=u_\mu T^{\mu\nu}u_\nu$, respectively, where $\Delta_{\mu\nu}=g_{\mu\nu}-u_\mu u_\nu$ represents the projection tensor. The velocity gradient should not be zero for the calculation of shear and bulk viscosities. In the local rest frame, $\Delta T^{00}=0$, so, only the spatial component of $\Delta T^{\mu\nu}$ is proportional to the velocity gradient. From eq. \eqref{em3}, the spatial component of $\Delta T^{\mu\nu}$ takes the following form, 
\be\label{em4}
\nonumber\Delta T^{ij} &=& \sum_f g_f\int\frac{d^3{\rm p}}{(2\pi)^3}\frac{p^i p^j}{\omega_f T}\left[\frac{\tau_f\left[1+(q-1)\beta(u_\alpha p^\alpha-\mu_f)\right]^{\frac{2-q}{q-1}}}{\left(\left[1+(q-1)\beta(u_\alpha p^\alpha-\mu_f)\right]^\frac{1}{q-1}+1\right)^2}\left\lbrace-\frac{p^kp^l}{2\omega_f}W_{kl}\right.\right. \\ && \left.\left.\nonumber+\left(\omega_f\left(\frac{\partial P}{\partial \varepsilon}\right)-\frac{\rm p^2}{3\omega_f}\right)\partial_l u^l-\frac{Tp^k}{\omega_f}\partial_k\left(\frac{\mu_f}{T}\right)+p^k\left(\frac{\partial_k P}{\varepsilon+P}-\frac{\partial_k T}{T} \right)\right\rbrace\right. \\ && \left.\nonumber+\frac{\tau_{\bar{f}}\left[1+(q-1)\beta(u_\alpha p^\alpha+\mu_f)\right]^{\frac{2-q}{q-1}}}{\left(\left[1+(q-1)\beta(u_\alpha p^\alpha+\mu_f)\right]^\frac{1}{q-1}+1\right)^2}\left\lbrace-\frac{p^kp^l}{2\omega_f}W_{kl}\right.\right. \\ && \left.\left.\nonumber +\left(\omega_f\left(\frac{\partial P}{\partial \varepsilon}\right)-\frac{\rm p^2}{3\omega_f}\right)\partial_l u^l+\frac{Tp^k}{\omega_f}\partial_k\left(\frac{\mu_f}{T}\right)+p^k\left(\frac{\partial_k P}{\varepsilon+P}-\frac{\partial_k T}{T} \right)\right\rbrace\right] \\ && \nonumber+g_g\int\frac{d^3{\rm p}}{(2\pi)^3}\frac{p^i p^j}{\omega_g T}\frac{\tau_g\left[1+(q-1)\beta u_\alpha p^\alpha\right]^{\frac{2-q}{q-1}}}{\left(\left[1+(q-1)\beta u_\alpha p^\alpha\right]^\frac{1}{q-1}-1\right)^2}\left[-\frac{p^kp^l}{2\omega_g}W_{kl}\right. \\ && \left.+\left(\omega_g\left(\frac{\partial P}{\partial \varepsilon}\right)-\frac{\rm p^2}{3\omega_g}\right)\partial_l u^l+p^k\left(\frac{\partial_k P}{\varepsilon+P}-\frac{\partial_k T}{T} \right)\right]
.\ee
In getting the above equation, we have used $\partial_k u_l=-\frac{1}{2}W_{kl}-\frac{1}{3}\delta_{kl}\partial_j u^j$ and $W_{kl}=\partial_k u_l+\partial_l u_k-\frac{2}{3}\delta_{kl}\partial_j u^j$. 

The shear and bulk viscosities are defined as the coefficients of the traceless and trace parts of the dissipative contribution of the energy-momentum tensor, respectively. In a first order theory, the 
spatial component of the nonequilibrium part of the energy-momentum tensor is given \cite{Lifshitz:BOOK'1981,Hosoya:NPB250'1985,Landau:BOOK'1987} by
\be\label{definition}
\Delta T^{ij}=-\eta W^{ij}-\zeta\delta^{ij}\partial_l u^l
.\ee
By comparing equations \eqref{em4} and \eqref{definition}, the shear and bulk viscosities are respectively obtained as
\begin{eqnarray}\label{iso.eta}
\nonumber\eta &=& \frac{\beta}{30\pi^2}\sum_f g_f \int d{\rm p}~\frac{{\rm p}^6}{\omega_f^2}\left[\frac{\tau_f\left[1+(q-1)\beta(u_\alpha p^\alpha-\mu_f)\right]^{\frac{2-q}{q-1}}}{\left(\left[1+(q-1)\beta(u_\alpha p^\alpha-\mu_f)\right]^\frac{1}{q-1}+1\right)^2}\right. \\ && \left.\nonumber+\frac{\tau_{\bar{f}}\left[1+(q-1)\beta(u_\alpha p^\alpha+\mu_f)\right]^{\frac{2-q}{q-1}}}{\left(\left[1+(q-1)\beta(u_\alpha p^\alpha+\mu_f)\right]^\frac{1}{q-1}+1\right)^2}\right] \\ && +\frac{\beta}{30\pi^2} g_g \int d{\rm p}~\frac{{\rm p}^6}{\omega_g^2}\frac{\tau_g\left[1+(q-1)\beta u_\alpha p^\alpha\right]^{\frac{2-q}{q-1}}}{\left(\left[1+(q-1)\beta u_\alpha p^\alpha\right]^\frac{1}{q-1}-1\right)^2}
~,\end{eqnarray}
\begin{eqnarray}\label{iso.zeta1}
\nonumber\zeta &=& \frac{1}{3}\sum_f g_f \int\frac{d^3{\rm p}}{(2\pi)^3}~\frac{{\rm p}^2}{\omega_f}\left[\frac{\left[1+(q-1)\beta(u_\alpha p^\alpha-\mu_f)\right]^{\frac{2-q}{q-1}}A_f}{\left(\left[1+(q-1)\beta(u_\alpha p^\alpha-\mu_f)\right]^\frac{1}{q-1}+1\right)^2}\right. \\ && \left.\nonumber+\frac{\left[1+(q-1)\beta(u_\alpha p^\alpha+\mu_f)\right]^{\frac{2-q}{q-1}}\bar{A}_f}{\left(\left[1+(q-1)\beta(u_\alpha p^\alpha+\mu_f)\right]^\frac{1}{q-1}+1\right)^2}\right] \\ && +\frac{1}{3}g_g \int\frac{d^3{\rm p}}{(2\pi)^3}~\frac{{\rm p}^2}{\omega_g}\frac{\left[1+(q-1)\beta u_\alpha p^\alpha\right]^{\frac{2-q}{q-1}}A_g}{\left(\left[1+(q-1)\beta u_\alpha p^\alpha\right]^\frac{1}{q-1}-1\right)^2}
~,\end{eqnarray}
where $A_f=\frac{\tau_f}{T}\left[\frac{{\rm p}^2}{3\omega_f}-\left(\frac{\partial P}{\partial \varepsilon}\right)\omega_f\right],$ $\bar{A}_f=\frac{\tau_{\bar{f}}}{T}\left[\frac{{\rm p}^2}{3\omega_f}-\left(\frac{\partial P}{\partial \varepsilon}\right)\omega_f\right],$ $A_g=\frac{\tau_g}{T}\left[\frac{{\rm p}^2}{3\omega_g}-\left(\frac{\partial P}{\partial \varepsilon}\right)\omega_g\right].$ 
In the local rest frame, in conjunction with the Landau-Lifshitz condition ($\Delta T^{00}=0$), the factors $A_f$, $\bar{A}_f$ and $A_g$ are replaced as $A_f\rightarrow A_f^\prime=A_f-b_f\omega_f$, $\bar{A}_f\rightarrow \bar{A}_f^\prime=\bar{A}_f-\bar{b}_f\omega_f$ and $A_g\rightarrow A_g^\prime=A_g-b_g\omega_g$, where $b_f$, $\bar{b}_f$ and $b_g$ are arbitrary constants and are related to the particle number and energy conservations for a thermal medium having asymmetry between the numbers of particles and antiparticles \cite{Chakraborty:PRC83'2011}. From the ``00" component of eq. \eqref{em3}, the Landau-Lifshitz conditions for $A_f$, $\bar{A}_f$ and $A_g$ are respectively written as
\begin{eqnarray}
&&\sum_f g_f\int\frac{d^3{\rm p}}{(2\pi)^3}\frac{\omega_f\left[1+(q-1)\beta(u_\alpha p^\alpha-\mu_f)\right]^{\frac{2-q}{q-1}}\left(A_f-b_f\omega_f\right)}{\left(\left[1+(q-1)\beta(u_\alpha p^\alpha-\mu_f)\right]^\frac{1}{q-1}+1\right)^2}=0 \label{A_i} ~,~ \\ 
&&\sum_f g_f\int\frac{d^3{\rm p}}{(2\pi)^3}\frac{\omega_f\left[1+(q-1)\beta(u_\alpha p^\alpha+\mu_f)\right]^{\frac{2-q}{q-1}}\left(\bar{A}_f-\bar{b}_f\omega_f\right)}{\left(\left[1+(q-1)\beta(u_\alpha p^\alpha+\mu_f)\right]^\frac{1}{q-1}+1\right)^2}=0 \label{A_i.1} ~,~ \\ 
&&g_g\int\frac{d^3{\rm p}}{(2\pi)^3}\frac{\omega_g\left[1+(q-1)\beta u_\alpha p^\alpha\right]^{\frac{2-q}{q-1}}\left(A_g-b_g\omega_g\right)}{\left(\left[1+(q-1)\beta u_\alpha p^\alpha\right]^\frac{1}{q-1}-1\right)^2}=0 \label{A_g}
~.\end{eqnarray}
The quantities $b_f$, $\bar{b}_f$ and $b_g$ can be calculated by solving equations \eqref{A_i}, \eqref{A_i.1} and \eqref{A_g}. By substituting $A_f\rightarrow A_f^\prime$, $\bar{A}_f\rightarrow \bar{A}_f^\prime$ and $A_g\rightarrow A_g^\prime$ in eq. \eqref{iso.zeta1} and then simplifying, we get the bulk 
viscosity as
\begin{eqnarray}\label{iso.zeta}
\nonumber\zeta &=& \frac{\beta}{18\pi^2}\sum_f g_f \int d{\rm p}~{\rm p}^2\left[\frac{{\rm p}^2}{\omega_f}-3\left(\frac{\partial P}{\partial \varepsilon}\right)\omega_f\right]^2\left[\frac{\tau_f\left[1+(q-1)\beta(u_\alpha p^\alpha-\mu_f)\right]^{\frac{2-q}{q-1}}}{\left(\left[1+(q-1)\beta(u_\alpha p^\alpha-\mu_f)\right]^\frac{1}{q-1}+1\right)^2}\right. \\ && \left.\nonumber+\frac{\tau_{\bar{f}}\left[1+(q-1)\beta(u_\alpha p^\alpha+\mu_f)\right]^{\frac{2-q}{q-1}}}{\left(\left[1+(q-1)\beta(u_\alpha p^\alpha+\mu_f)\right]^\frac{1}{q-1}+1\right)^2}\right] \\ && +\frac{\beta}{18\pi^2}g_g\int d{\rm p}~{\rm p}^2\left[\frac{{\rm p}^2}{\omega_g}-3\left(\frac{\partial P}{\partial \varepsilon}\right)\omega_g\right]^2\frac{\tau_g\left[1+(q-1)\beta u_\alpha p^\alpha\right]^{\frac{2-q}{q-1}}}{\left(\left[1+(q-1)\beta u_\alpha p^\alpha\right]^\frac{1}{q-1}-1\right)^2}
~.\end{eqnarray}

\subsection{Nonextensive hot QCD medium in the presence of a strong magnetic field}
The presence of a strong magnetic field markedly affects the dynamics of charged particles in the medium. They tend to move along the direction of magnetic field (say, z or $3$-direction). Thus, the quark momentum 
becomes bifurcated into the transverse ($p_T$) and longitudinal ($p_L$) components. As a result, the dispersion relation of the quark of $f$th flavor gets modified into 
\begin{eqnarray}\label{dispersion relation}
\omega_{f,n}(p_L)=\sqrt{p_L^2+2n\left|q_fB\right|+m_f^2}
~.\end{eqnarray}
Here, the transverse motion is quantized and $n=0$, $1$, $2$, $\cdots$ represent different Landau levels. In the strong magnetic field (SMF) limit ($|q_fB| \gg T^2$), the transitions of quarks to $n>1$ levels are forbidden as the energy gap between the levels is of the order $\sim\mathcal{O}(\sqrt{|q_fB|})$. Hence, the quarks stay only in the lowest Landau levels ($n=0$). In the strong magnetic field regime, the nonextensive fermion distribution function is written as
\be\label{A.D.F.(eB)}
&&f^B=\frac{1}{\left[1+(q-1)\beta(u^\alpha\tilde{p}_\alpha\mp\mu_f)\right]^{\frac{1}{q-1}}+1}
~,\ee
where `$-$' sign is for quark case ($f^B_q$), `$+$' sign is for antiquark case ($\bar{f}^B_q$), $\tilde{p}_\alpha\equiv\left(\omega_f,p_3\right)$ and $\omega_f=\sqrt{p_3^2+m_f^2}$. But, the gluons being electrically uncharged particles do not get affected by the presence of magnetic field, so, the form of the nonextensive gluon distribution function remains unaltered. 

In the strong magnetic field regime, the energy-momentum tensor for the nonequilibrium system is defined as
\be
\tilde{T^\prime}^{\mu\nu}=\sum_f \frac{g_f|q_fB|}{4\pi^2}\int d p_3 ~ \frac{\tilde{p}^\mu\tilde{p}^\nu}{\omega_f}\left(f_q^\prime+\bar{f}_q^\prime\right)
,\ee
where $f_q^\prime=f^B_q+\delta f_q$ and $\bar{f}_q^\prime=\bar{f}^B_q+\delta \bar{f}_q$. In the above equation, one can notice the modified (integration) phase factor due to the strong magnetic field, {\em i.e.} $\int\frac{d^3{\rm p}}{(2\pi)^3}=\frac{|q_fB|}{4\pi^2}\int dp_3$. In the aforesaid regime, the dissipative part of the energy-momentum tensor is written as
\be\label{emb1}
\Delta\tilde{T}^{\mu\nu}=\sum_f \frac{g_f|q_fB|}{4\pi^2}\int d p_3 ~ \frac{\tilde{p}^\mu\tilde{p}^\nu}{\omega_f}\left(\delta f_q+\delta \bar{f}_q\right)
.\ee
Here, $\tilde{p}^\mu=(p^0,0,0,p^3)$ as per the SMF limit. One can find the infinitesimal shifts, $\delta f_q$ and $\delta \bar{f}_q$ by solving the relativistic Boltzmann transport equations for quarks and antiquarks 
in the relaxation time approximation in the SMF limit as $\tilde{p}^\mu\partial_\mu f_q^\prime(x,p)=-\frac{\tilde{p}_\nu u^\nu}{\tau^B_f}\delta f_q ~,$ $\tilde{p}^\mu\partial_\mu \bar{f}_q^\prime(x,p)=-\frac{\tilde{p}_\nu u^\nu}{\tau^B_{\bar{f}}}\delta \bar{f}_q ~,$ where $\tau^B_{f(\bar{f})}$ denotes the relaxation time in the strong magnetic field limit and is expressed \cite{Hattori:PRD95'2017} as
\begin{eqnarray}
\tau^B_{f(\bar{f})}=\frac{\omega_f\left(e^{\beta\omega_f}-1\right)}{\alpha_sC_2m_f^2\left(e^{\beta\omega_f}+1\right)}\frac{1}{\int dp^\prime_3\frac{1}{\omega^\prime_f\left(e^{\beta\omega^\prime_f}+1\right)}}
~.\end{eqnarray}
Here, $C_2$ represents the Casimir factor. Now eq. \eqref{emb1} becomes
\be\label{emb2}
\Delta\tilde{T}^{\mu\nu}=-\sum_f \frac{g_f|q_fB|}{4\pi^2}\int d p_3 ~ \frac{\tilde{p}^\mu\tilde{p}^\nu}{\tilde{p}_\nu u^\nu \omega_f}\left(\tau^B_f \tilde{p}^\mu\partial_\mu f_q^\prime+\tau^B_{\bar{f}} \tilde{p}^\mu\partial_\mu \bar{f}_q^\prime\right)
.\ee
The partial derivatives of the nonextensive quark and antiquark distribution functions in eq. \eqref{emb2} in the presence of a strong magnetic field are respectively determined as
\begin{eqnarray}
\nonumber\partial_\mu f_q^\prime=\frac{\beta\left[1+(q-1)\beta(u^\alpha\tilde{p}_\alpha-\mu_f)\right]^{\frac{2-q}{q-1}}}{\left(\left[1+(q-1)\beta(u^\alpha\tilde{p}_\alpha-\mu_f)\right]^\frac{1}{q-1}+1\right)^2}\left[u_\alpha \tilde{p}^\alpha u_\mu\frac{DT}{T}+u_\alpha \tilde{p}^\alpha\frac{\nabla_\mu T}{T}-u_\mu \tilde{p}^\alpha Du_\alpha\right. \\ \left.-\tilde{p}^\alpha\nabla_\mu u_\alpha+T\partial_\mu\left(\frac{\mu_f}{T}\right)\right]
,\end{eqnarray}
\begin{eqnarray}
\nonumber\partial_\mu \bar{f}_q^\prime=\frac{\beta\left[1+(q-1)\beta(u^\alpha\tilde{p}_\alpha+\mu_f)\right]^{\frac{2-q}{q-1}}}{\left(\left[1+(q-1)\beta(u^\alpha\tilde{p}_\alpha+\mu_f)\right]^\frac{1}{q-1}+1\right)^2}\left[u_\alpha \tilde{p}^\alpha u_\mu\frac{DT}{T}+u_\alpha \tilde{p}^\alpha\frac{\nabla_\mu T}{T}-u_\mu \tilde{p}^\alpha Du_\alpha\right. \\ \left.-\tilde{p}^\alpha\nabla_\mu u_\alpha-T\partial_\mu\left(\frac{\mu_f}{T}\right)\right]
.\end{eqnarray}
Substituting the values of $\partial_\mu f_q^\prime$ and $\partial_\mu \bar{f}_q^\prime$ in eq. \eqref{emb2} and then simplifying, we get 
\be\label{emb2.1}
\nonumber\Delta \tilde{T}^{\mu\nu} &=& \sum_f \frac{g_f|q_fB|}{4\pi^2}\int d p_3 ~ \frac{\tilde{p}^\mu \tilde{p}^\nu}{\omega_f T}\left[\frac{\tau_f^B\left[1+(q-1)\beta(u^\alpha\tilde{p}_\alpha-\mu_f)\right]^{\frac{2-q}{q-1}}}{\left(\left[1+(q-1)\beta(u^\alpha\tilde{p}_\alpha-\mu_f)\right]^\frac{1}{q-1}+1\right)^2}\left\lbrace\omega_f\left(\frac{\partial P}{\partial \varepsilon}\right)\nabla_\alpha u^\alpha\right.\right. \\ && \left.\left.\nonumber+\tilde{p}^\alpha\left(\frac{\nabla_\alpha P}{\varepsilon+P}-\frac{\nabla_\alpha T}{T}\right)-\frac{T\tilde{p}^\alpha}{\omega_f}\partial_\alpha\left(\frac{\mu_f}{T}\right)+\frac{\tilde{p}^\alpha \tilde{p}^\beta}{\omega_f}\nabla_\alpha u_\beta\right\rbrace\right. \\ && \left.\nonumber+\frac{\tau^B_{\bar{f}}\left[1+(q-1)\beta(u^\alpha\tilde{p}_\alpha+\mu_f)\right]^{\frac{2-q}{q-1}}}{\left(\left[1+(q-1)\beta(u^\alpha\tilde{p}_\alpha+\mu_f)\right]^\frac{1}{q-1}+1\right)^2}\left\lbrace\omega_f\left(\frac{\partial P}{\partial \varepsilon}\right)\nabla_\alpha u^\alpha\right.\right. \\ && \left.\left.+\tilde{p}^\alpha\left(\frac{\nabla_\alpha P}{\varepsilon+P}-\frac{\nabla_\alpha T}{T}\right)+\frac{T\tilde{p}^\alpha}{\omega_f}\partial_\alpha\left(\frac{\mu_f}{T}\right)+\frac{\tilde{p}^\alpha \tilde{p}^\beta}{\omega_f}\nabla_\alpha u_\beta\right\rbrace\right]
.\ee
The spatial or longitudinal component of $\Delta \tilde{T}^{\mu\nu}$ is written as
\be\label{emb3}
\nonumber\Delta \tilde{T}^{ij} &=& \sum_f \frac{g_f|q_fB|}{4\pi^2}\int d p_3 ~ \frac{\tilde{p}^i \tilde{p}^j}{\omega_f T}\left[\frac{\tau_f^B\left[1+(q-1)\beta(u^\alpha\tilde{p}_\alpha-\mu_f)\right]^{\frac{2-q}{q-1}}}{\left(\left[1+(q-1)\beta(u^\alpha\tilde{p}_\alpha-\mu_f)\right]^\frac{1}{q-1}+1\right)^2}\left\lbrace-\frac{\tilde{p}^k\tilde{p}^l}{2\omega_f}W_{kl}\right.\right. \\ && \left.\left.\nonumber+\left(\omega_f\left(\frac{\partial P}{\partial \varepsilon}\right)-\frac{p_3^2}{3\omega_f}\right)\partial_l u^l-\frac{T\tilde{p}^k}{\omega_f}\partial_k\left(\frac{\mu_f}{T}\right)+\tilde{p}^k\left(\frac{\partial_k P}{\varepsilon+P}-\frac{\partial_k T}{T} \right)\right\rbrace\right. \\ && \left.\nonumber+\frac{\tau^B_{\bar{f}}\left[1+(q-1)\beta(u^\alpha\tilde{p}_\alpha+\mu_f)\right]^{\frac{2-q}{q-1}}}{\left(\left[1+(q-1)\beta(u^\alpha\tilde{p}_\alpha+\mu_f)\right]^\frac{1}{q-1}+1\right)^2}\left\lbrace-\frac{\tilde{p}^k\tilde{p}^l}{2\omega_f}W_{kl}\right.\right. \\ && \left.\left. +\left(\omega_f\left(\frac{\partial P}{\partial \varepsilon}\right)-\frac{p_3^2}{3\omega_f}\right)\partial_l u^l+\frac{T\tilde{p}^k}{\omega_f}\partial_k\left(\frac{\mu_f}{T}\right)+\tilde{p}^k\left(\frac{\partial_k P}{\varepsilon+P}-\frac{\partial_k T}{T} \right)\right\rbrace\right]
.\ee
In the SMF limit, thermodynamic quantities like pressure and energy density are evaluated from the energy-momentum tensor through the relations, $P=-\Delta^\parallel_{\mu\nu}\tilde{T}^{\mu\nu}$ and $\varepsilon=u_\mu\tilde{T}^{\mu\nu}u_\nu$, respectively. Here, $\Delta^\parallel_{\mu\nu}$ is the longitudinal projection tensor, $\Delta^\parallel_{\mu\nu}=g^\parallel_{\mu\nu}-u_\mu u_\nu$, with $g^\parallel_{\mu\nu}$=${\rm{diag}}(1,0,0,-1)$ being the metric tensor. As compared to the zero magnetic field case, where there exist only two ordinary viscous coefficients ($\eta$ and $\zeta$ in eq. \eqref{definition}), the medium at finite magnetic field possesses seven viscous coefficients, namely, five shear viscous coefficients, $\eta$, $\eta_1$, $\eta_2$, $\eta_3$ and $\eta_4$, one bulk viscous coefficient, $\zeta$ and a cross-effect between the ordinary and bulk viscosities, $\zeta_1$. Thus, for an arbitrary magnetic field (${\bf B}$ along a direction, $\bf b=\frac{\bf B}{\rm B}$), the viscous tensor takes the following form \cite{Lifshitz:BOOK'1981}, 
\begin{align}\label{Form1}
\pi_{ij} =& 2 \eta\left(V_{ij}-\frac{1}{3}
\delta_{ij} \nabla \cdot \mathbf V \right) +\zeta 
\delta_{ij} \nabla \cdot \mathbf V\nonumber \\
& +\eta_1\left(2V_{ij}-\delta_{ij}\nabla\cdot\mathbf{V} +
\delta_{ij}V_{kl}b_k b_l-2V_{ik}
b_k b_j-2V_{jk}b_k b_i+b_i b_j\nabla
\cdot\mathbf{V}+b_i b_j V_{kl}b_k b_l\right) 
\nonumber \\
& +2\eta_2\left(V_{ik}b_k b_j+V_{jk}b_k b_i-2b_i b_j V_{kl}b_k b_l
\right)\nonumber \\
& +\eta_3\left(V_{ik}b_{jk}+V_{jk}
b_{ik}-V_{kl}b_{ik}b_j b_l-
V_{kl}b_{jk}b_i b_l\right) \nonumber \\
& +2\eta_4\left(V_{kl}b_{ik}b_j b_l +
V_{kl}b_{jk}b_i b_l\right)\nonumber \\
& +\zeta_1\left(\delta_{ij}V_{kl}b_k b_l+
b_i b_j\nabla\cdot\mathbf{V}\right)
,\end{align}
where $b_{ij}=\epsilon_{ijk}b_k$ and $V_{ij}=\frac{1}{2}
\left(\frac{\partial V_i}{\partial x_j}
+\frac{\partial V_j}{\partial x_i}\right)$. In eq. \eqref{Form1}, $\eta$, $\eta_1$, $\eta_2$, $\eta_3$ and $\eta_4$ are the coefficients of the traceless part of $\pi_{ij}$, and $\zeta$ and $\zeta_1$ are the coefficients of the finite trace part of $\pi_{ij}$. In this equation, $\eta$ and $\zeta$ are known as the ordinary viscosity coefficients, because the terms containing them are same as the terms at $B=0$ case. However, for a plasma in the strong magnetic field regime, $\eta_1, \eta_2, \eta_3$, $\eta_4$ and $\zeta_1$ coefficients vanish and eq. \eqref{Form1} takes a much simpler form through the replacement of the $\eta$-term in the above equation by $\eta \left(3 b_i b_j - \delta_{ij}\right) \left(b_k b_l V_{kl} -\frac{1}{3} \nabla \cdot V\right)$ (for a detailed discussion, see references \cite{Lifshitz:BOOK'1981,Rath:PRD102'2020}). Thus, in the SMF limit, the viscous tensor gets generalized into the relativistic energy-momentum tensor, ${\tilde{T}}^{\mu \nu}$ \cite{Lifshitz:BOOK'1981,Ofengeim:EPL112'2015}, whose dissipative part is defined as
\begin{equation}\label{FORM (1)}
\Delta\tilde{T}^{\mu\nu} = -\eta\left(\frac{\partial{u}^\mu}{\partial\tilde{x}_\nu}+\frac{\partial{u}^\nu}{\partial\tilde{x}_\mu}-{u}^\nu{u}_\lambda\frac{\partial{u}^\mu}{\partial\tilde{x}_\lambda}-{u}^\mu{u}_\lambda\frac{\partial{u}^\nu}{\partial\tilde{x}_\lambda}-\frac{2}{3}\Delta_\parallel^{\mu\nu}\frac{\partial{u}^\lambda}{\partial\tilde{x}^\lambda}\right)-\zeta\Delta_\parallel^{\mu\nu}\frac{\partial{u}^\lambda}{\partial\tilde{x}^\lambda}
~.\end{equation}
Here, $\tilde{x}^\mu=(x^0,0,0,x^3)$. In the local rest frame, the spatial component of $\Delta\tilde{T}^{\mu\nu}$ (eq. \eqref{FORM (1)}) is written \cite{Lifshitz:BOOK'1981,Ofengeim:EPL112'2015} as
\begin{align}\label{definition(eb)}
\nonumber\Delta\tilde{T}^{ij} =& -\eta\left(\frac{\partial{u}^i}{\partial\tilde{x}_j}+\frac{\partial{u}^j}{\partial\tilde{x}_i}-\frac{2}{3}\delta^{ij}\frac{\partial{u}^l}{\partial\tilde{x}^l}\right)
-\zeta\delta^{ij}\frac{\partial{u}^l}{\partial\tilde{x}^l} \\ =& \nonumber -\eta\left(\partial^i{u}^j+\partial^j{u}^i-\frac{2}{3}\delta^{ij}\partial_l{u}^l\right)
-\zeta\delta^{ij}\partial_l{u}^l \\ =& -\eta{W}^{ij}-\zeta\delta^{ij}\partial_l{u}^l
.\end{align}
Through the comparison of eq. \eqref{emb3} and eq. \eqref{definition(eb)}, one can obtain the charged particle (quarks and antiquarks) contribution to the shear viscosity in a nonextensive medium at strong magnetic field as
\be\label{A.S.V.q(eB)}
\nonumber\eta_q &=& \frac{\beta}{8\pi^2}\sum_f g_f ~ |q_fB|\int dp_3~\frac{p_3^4}{\omega_f^2}\left[\frac{\tau_f^B\left[1+(q-1)\beta(u^\alpha\tilde{p}_\alpha-\mu_f)\right]^{\frac{2-q}{q-1}}}{\left(\left[1+(q-1)\beta(u^\alpha\tilde{p}_\alpha-\mu_f)\right]^\frac{1}{q-1}+1\right)^2}\right. \\ && \left.+\frac{\tau^B_{\bar{f}}\left[1+(q-1)\beta(u^\alpha\tilde{p}_\alpha+\mu_f)\right]^{\frac{2-q}{q-1}}}{\left(\left[1+(q-1)\beta(u^\alpha\tilde{p}_\alpha+\mu_f)\right]^\frac{1}{q-1}+1\right)^2}\right]
.\ee
Since gluons are not affected by the magnetic field, one can add the gluon part to the magnetized charged particle part to obtain the total shear viscosity, {\em i.e.}, 
\be\label{A.S.V.(eB)}
\nonumber\eta &=& \frac{\beta}{8\pi^2}\sum_f g_f ~ |q_fB|\int dp_3~\frac{p_3^4}{\omega_f^2}\left[\frac{\tau_f^B\left[1+(q-1)\beta(u^\alpha\tilde{p}_\alpha-\mu_f)\right]^{\frac{2-q}{q-1}}}{\left(\left[1+(q-1)\beta(u^\alpha\tilde{p}_\alpha-\mu_f)\right]^\frac{1}{q-1}+1\right)^2}\right. \\ && \left.\nonumber+\frac{\tau^B_{\bar{f}}\left[1+(q-1)\beta(u^\alpha\tilde{p}_\alpha+\mu_f)\right]^{\frac{2-q}{q-1}}}{\left(\left[1+(q-1)\beta(u^\alpha\tilde{p}_\alpha+\mu_f)\right]^\frac{1}{q-1}+1\right)^2}\right] \\ && +\frac{\beta}{30\pi^2} g_g \int d{\rm p}~\frac{{\rm p}^6}{\omega_g^2}\frac{\tau_g\left[1+(q-1)\beta u_\alpha p^\alpha\right]^{\frac{2-q}{q-1}}}{\left(\left[1+(q-1)\beta u_\alpha p^\alpha\right]^\frac{1}{q-1}-1\right)^2}
.\ee
In the similar way, one can get the charged particle (quarks and antiquarks) contribution to the bulk viscosity by comparing eq. \eqref{emb3} and eq. \eqref{definition(eb)}, 
\be\label{A.B.V.q(eB)}
\nonumber\zeta_q &=& \sum_f \frac{g_f|q_fB|}{4\pi^2}\int d p_3~\frac{p_3^2}{\omega_f}\left[\frac{\left[1+(q-1)\beta(u^\alpha\tilde{p}_\alpha-\mu_f)\right]^{\frac{2-q}{q-1}}A_f}{\left(\left[1+(q-1)\beta(u^\alpha\tilde{p}_\alpha-\mu_f)\right]^\frac{1}{q-1}+1\right)^2} \right. \\ && \left.+\frac{\left[1+(q-1)\beta(u^\alpha\tilde{p}_\alpha+\mu_f)\right]^{\frac{2-q}{q-1}}\bar{A}_f}{\left(\left[1+(q-1)\beta(u^\alpha\tilde{p}_\alpha+\mu_f)\right]^\frac{1}{q-1}+1\right)^2}\right]
,\ee
where $A_f=\frac{\tau_f^B}{T}\left[\frac{p_3^2}{3\omega_f}-\left(\frac{\partial P}{\partial \varepsilon}\right)\omega_f\right],$ $\bar{A}_f=\frac{\tau_{\bar{f}}^B}{T}\left[\frac{p_3^2}{3\omega_f}-\left(\frac{\partial P}{\partial \varepsilon}\right)\omega_f\right].$ Using the Landau-Lifshitz condition in the calculation of the bulk viscosity and then simplifying, we finally get
\be\label{A.B.V.q1(eB)}
\nonumber\zeta_q &=& \frac{\beta}{12\pi^2}\sum_f g_f|q_fB|\int d p_3~\left[\frac{p_3^2}{\omega_f}-3\left(\frac{\partial P}{\partial\varepsilon}\right)\omega_f\right]^2\left[\frac{\tau_f^B\left[1+(q-1)\beta(u^\alpha\tilde{p}_\alpha-\mu_f)\right]^{\frac{2-q}{q-1}}}{\left(\left[1+(q-1)\beta(u^\alpha\tilde{p}_\alpha-\mu_f)\right]^\frac{1}{q-1}+1\right)^2}\right. \\ && \left.+\frac{\tau^B_{\bar{f}}\left[1+(q-1)\beta(u^\alpha\tilde{p}_\alpha+\mu_f)\right]^{\frac{2-q}{q-1}}}{\left(\left[1+(q-1)\beta(u^\alpha\tilde{p}_\alpha+\mu_f)\right]^\frac{1}{q-1}+1\right)^2}\right]
.\ee
The total bulk viscosity is obtained by adding the gluon part to the magnetized charged particle part as
\be\label{A.B.V.(eB)}
\nonumber\zeta &=& \frac{\beta}{12\pi^2}\sum_f g_f|q_fB|\int d p_3~\left[\frac{p_3^2}{\omega_f}-3\left(\frac{\partial P}{\partial\varepsilon}\right)\omega_f\right]^2\left[\frac{\tau_f^B\left[1+(q-1)\beta(u^\alpha\tilde{p}_\alpha-\mu_f)\right]^{\frac{2-q}{q-1}}}{\left(\left[1+(q-1)\beta(u^\alpha\tilde{p}_\alpha-\mu_f)\right]^\frac{1}{q-1}+1\right)^2}\right. \\ && \left.\nonumber+\frac{\tau^B_{\bar{f}}\left[1+(q-1)\beta(u^\alpha\tilde{p}_\alpha+\mu_f)\right]^{\frac{2-q}{q-1}}}{\left(\left[1+(q-1)\beta(u^\alpha\tilde{p}_\alpha+\mu_f)\right]^\frac{1}{q-1}+1\right)^2}\right] \\ && +\frac{\beta}{18\pi^2}g_g\int d{\rm p}~{\rm p}^2\left[\frac{{\rm p}^2}{\omega_g}-3\left(\frac{\partial P}{\partial \varepsilon}\right)\omega_g\right]^2\frac{\tau_g\left[1+(q-1)\beta u_\alpha p^\alpha\right]^{\frac{2-q}{q-1}}}{\left(\left[1+(q-1)\beta u_\alpha p^\alpha\right]^\frac{1}{q-1}-1\right)^2}
.\ee

\section{Results and discussions}
In this section, the results on the response of the nonextensivity on the viscous properties of hot QCD medium are discussed. The two scenarios considered for the discussion are (i) the nonextensive medium at zero magnetic field and (ii) the nonextensive medium at strong magnetic field. The study also incorporates the quasiparticle model of partons where QGP is treated as a medium consisting of thermally 
massive noninteracting quasiparticles. The partons gain thermal masses due to their interactions with the surrounding medium. For a thermal QCD medium in the absence of magnetic field, the available energy scales are associated with the temperature, chemical potential and quark mass. In the QGP phase consisting of light quarks and gluons, the temperature is high enough to become the largest energy scale. In this high temperature regime, the divergences encountered in the calculation of QCD thermodynamic observables, transport coefficients and amplitudes are cured by applying the effective theories, one of which is the hard thermal loop (HTL) perturbation theory. In this theory, the loop momentum is of the order of $T$, {\em i.e.} the hard scale and the next scale is the soft scale of the order of $gT$ ($g$ is the coupling constant). When the thermal medium is exposed to a strong magnetic field, $\sqrt{eB}$ emerges as a new energy scale of the system in addition to $T$ and $\mu$. Depending on the strength of the magnetic field compared to the temperature, the thermal medium may be considered as weakly magnetized or strongly magnetized. In particular, for a strongly magnetized thermal medium where $\sqrt{eB}\gg{T}$ and $\sqrt{eB}\gg{\mu}$, magnetic field plays the role of largest energy scale. In this case, the hard scales are different to the quark and gluon degrees of freedom, unlike the single hard scale for both degrees of freedom in thermal medium in the absence of magnetic field. The magnetic field and the temperature can be treated as hard scales for quark and gluon degrees of freedom in thermal quantum chromodynamics in a strong magnetic field, respectively. In addition, $g\sqrt{eB}$ arises as a new soft scale and the self-energies of $\mathcal{O}(g^2eB)$ may be considered as small corrections in the momentum regime, $p\gg g\sqrt{eB}$. Thus, there exists a hierarchy of energy scale in thermal QCD in a strong magnetic field, such as $gT \ll g\sqrt{eB} \ll T \ll \sqrt{eB}$. The quark loops are mainly affected by the strong magnetic field, whereas the gluon loops remain almost unaffected. So, in the strong magnetic field regime also, to alleviate the divergences associated with the gluon loops, the same resummation technique as in pure thermal medium, {\em i.e.} the HTL perturbation theory could be used. On the other hand, for quark loops in the strong magnetic field limit, the upper limit of the loop momentum is replaced by the magnetic field ($\sqrt{eB}$), because for quarks the dominant scale in this limit is the magnetic 
field, like the temperature acts as the dominant scale for gluons as well as for the case in the absence 
of magnetic field. 

For a QCD medium in the absence of magnetic field, the thermal mass (squared) of quark of $f$th flavor is written \cite{Braaten:PRD45'1992,Peshier:PRD66'2002} as
\be\label{Q.P.M.}
m_{fT}^2=\frac{g^{\prime2}T^2}{6}\left(1+\frac{\mu_f^2}{\pi^2T^2}\right)
.\ee
Here, $g^\prime$ represents the one-loop running coupling at zero magnetic field with the following \cite{Kapusta:BOOK'2006,Mustafa:EPJST232'2023} form, 
\begin{eqnarray}
g^{\prime2}=4\pi\alpha_s^\prime=\frac{48\pi^2}{\left(11N_c-2N_f\right)\ln\left({\Lambda^2}/{\Lambda_{\rm\overline{MS}}^2}\right)}
~,\end{eqnarray}
where $\Lambda_{\rm\overline{MS}}=0.176$ GeV, $\Lambda=2\pi\sqrt{T^2+\mu_f^2/\pi^2}$ for electrically charged particles (quarks and antiquarks) and $\Lambda=2 \pi T$ for gluons. In the semiclassical transport theory, the thermal mass (squared) of gluon is given \cite{Kelly:PRD50'1994,Litim:PR364'2002} by
\be\label{Q.P.M.(definition of gluon mass)}
\nonumber m_{gT}^2 &=& \frac{g^{\prime2}N_c}{2\pi^2T}\int d{\rm k}~\frac{{\rm k}^3}{\omega_g}\frac{e^{\beta\omega_g}}{\left(e^{\beta\omega_g}-1\right)^2}+\frac{g^{\prime2}N_f}{4\pi^2T}\int d{\rm k}~\frac{{\rm k}^3}{\omega_f}\left[\frac{e^{\beta(\omega_f-\mu_f)}}{\left(e^{\beta(\omega_f-\mu_f)}+1\right)^2}\right. \\ && \left.+\frac{e^{\beta(\omega_f+\mu_f)}}{\left(e^{\beta(\omega_f+\mu_f)}+1\right)^2}\right]
.\ee
The integrals in the above equation can be calculated using the hard thermal loop approximation and the simplified form of $m_{gT}^2$ is written \cite{Peshier:PRD66'2002,Blaizot:PRD72'2005,Berrehrah:PRC89'2014} as
\be\label{Q.P.M.(Gluon mass)}
m_{gT}^2=\frac{g^{\prime2}T^2}{6}\left(N_c+\frac{N_f}{2}+\frac{3}{2\pi^2T^2}\sum_f\mu_f^2\right)
.\ee
In the presence of a strong magnetic field, the form of the thermal gluon 
mass (squared) gets changed \cite{Rath:EPJC81'2021} to 
\be\label{Q.P.M.eB(Gluon mass)}
m_{gT,B}^2=\frac{g^{\prime2}T^2N_c}{6}+\frac{g^2}{8\pi^2T}\sum_f|q_fB|\int dk_z~\frac{k_z}{\omega_f}\left[\frac{e^{\beta(\omega_f-\mu_f)}}{\left(e^{\beta(\omega_f-\mu_f)}+1\right)^2}+\frac{e^{\beta(\omega_f+\mu_f)}}{\left(e^{\beta(\omega_f+\mu_f)}+1\right)^2}\right]
.\ee
Here, $g$ denotes the one-loop running coupling in the strong magnetic field limit and it runs exclusively with the magnetic field. It is given \cite{Ferrer:PRD91'2015} by
\begin{equation}
g^2=4\pi\alpha_s=\frac{4\pi}{{\alpha_s^0(\mu_0)}^{-1}+\frac{11N_c}{12\pi}
\ln\left(\frac{\Lambda_{QCD}^2+M^2_B}{\mu_0^2}\right)+\frac{1}{3\pi}\sum_f \frac{|q_f B|}{\tau}}
~,\end{equation}
where 
\begin{equation}
\alpha_s^0(\mu_0) = \frac{12\pi}
{11N_c\ln\left(\frac{\mu_0^2+M^2_B}{\Lambda_V^2}\right)}
\end{equation}
and $M_B$ ($=\sqrt{2\pi\tau}\approx1$ GeV) represents an infrared mass which is interpreted as the 
ground state mass of two gluons connected by a fundamental string, with the 
string tension, $\tau=0.18 ~ {\rm{GeV}}^2$. For this numerical value of $M_B$ as input, the 
values of factors $\Lambda_V=0.385$ GeV and $\mu_0=1.1$ GeV are chosen to make $\alpha_s$ small, 
{\em i.e.} $\alpha_s<1$ \cite{Ferrer:PRD91'2015,Simonov:PAN58'1995,Andreichikov:PRL110'2013}. The thermal mass of gluon is also related to the Debye mass as $m_D^2=2m_{gT,B}^2$. Thus, one can obtain the Debye mass in a strong magnetic field from eq. \eqref{Q.P.M.eB(Gluon mass)}. This Debye mass is also comparable to the Debye mass obtained in ref. \cite{Bandyopadhyay:PRD100'2019} within the strong magnetic field regime. We have found that, in the limits $m_f\rightarrow 0$ and $\mu_f\rightarrow 0$, {\em i.e.} if one neglects the masses and chemical potentials of $u$ and $d$ quarks, the Debye mass in this work becomes exactly equal to the Debye mass (eq. (57)) calculated in the strong magnetic field limit in ref. \cite{Bandyopadhyay:PRD100'2019}. However 
for the three flavor case, it is not possible to compare these results at the equal base, because one cannot neglect the mass of $s$ quark. In addition, the thermal quark mass (eq. \eqref{Q.P.M.}) also gets altered by the strong magnetic field. It can be determined by taking $p_0=0, p_z\rightarrow 0$ limit of the effective quark propagator through the Schwinger-Dyson equation in the presence of a strong magnetic field. Its form is given \cite{Rath:EPJC80'2020} by
\begin{eqnarray}\label{Mass}
m_{fT,B}^2=\frac{g^2|q_fB|}{3\pi^2}\left[\frac{\pi T}{2m_f}-\ln(2)+\frac{7\mu_f^2\zeta(3)}{8\pi^2T^2}-\frac{31\mu_f^4\zeta(5)}{32\pi^4T^4}\right]
.\end{eqnarray}
We note that, all flavors are assigned the same chemical potential ($\mu_f=\mu$). We further note that the thermal mass of quark in the strong magnetic field limit \eqref{Mass} was obtained 
from the Schwinger-Dyson equation for the one-loop case \cite{Rath:EPJC80'2020}. For the pure thermal medium (absence of magnetic field), the one-loop self-energy for massless case (which is true even for finite mass case if the temperature is much higher than the mass) is $\sim {\cal O}(g^2T^2)$ and the average of the square of momentum at the hard scale ($T$) is $\sim {\cal O} (T^2)$, thus the one-loop self-energy can be termed as a correction because $T^2 \gg g^2T^2$. For the thermal medium in the presence of a strong magnetic field, the one-loop self-energy is $\sim {\cal O} (g^2|q_fB|)$ and the average of the square of momentum at the hard scale ($\sqrt{|q_fB|}$) due to the strong magnetic field limit ($|q_fB| \gg T^2$) is $\sim {\cal O}(|q_fB|)$, so the one-loop self-energy even in the presence of a strong magnetic field still acts as a correction. To be precise, the analogous part $\frac{g^2|q_fB|}{3\pi^2}$ (in a strong magnetic field) is larger than its counterpart $\frac{g^{\prime2}T^2}{6}$ (in the absence of magnetic field) if the magnetic field is sufficiently large. However, one of the terms in the full expression of the one-loop self-energy (hence the (square) thermal mass), {\em i.e.} $\frac{\pi T}{2m_f}$ looks divergent in case of $u$ quark or $d$ 
quark. This term arises precisely in the (one-dimensional) momentum integration (due to the dimensional reduction by the strong magnetic field), where the divergent-like behavior in the lower limit is tamed by the current quark masses. For the comparison, this dimensional reduction does not arise in pure thermal medium, hence this kind of term does not appear in the thermal mass of quark in the absence of magnetic field. In the quasiparticle model, the $T$ and $\mu$-dependent quark \eqref{Q.P.M.} and gluon \eqref{Q.P.M.(Gluon mass)} masses are used for the QCD medium at finite temperature and chemical potential, and the $T$, $\mu$ and $eB$-dependent quark \eqref{Mass} and gluon \eqref{Q.P.M.eB(Gluon mass)} masses are used in the presence of a strong magnetic field. In the calculation, we have chosen a specific range of temperature, magnetic field and chemical potential in accordance with the strong magnetic field limit ($eB \gg T^2$, $eB \gg \mu^2$). Thus, while computing different transport coefficients and observables as functions 
of temperature up to $T=0.4$ GeV, we have fixed the magnetic field at $eB=15 m_\pi^2$ and the chemical potential at $\mu=0.06$ GeV, with the conversion factor, $1 m_\pi^2 \sim 0.02 ~ {\rm GeV}^2$. 

\begin{figure}[]
\begin{center}
\begin{tabular}{c c}
\includegraphics[width=6.3cm]{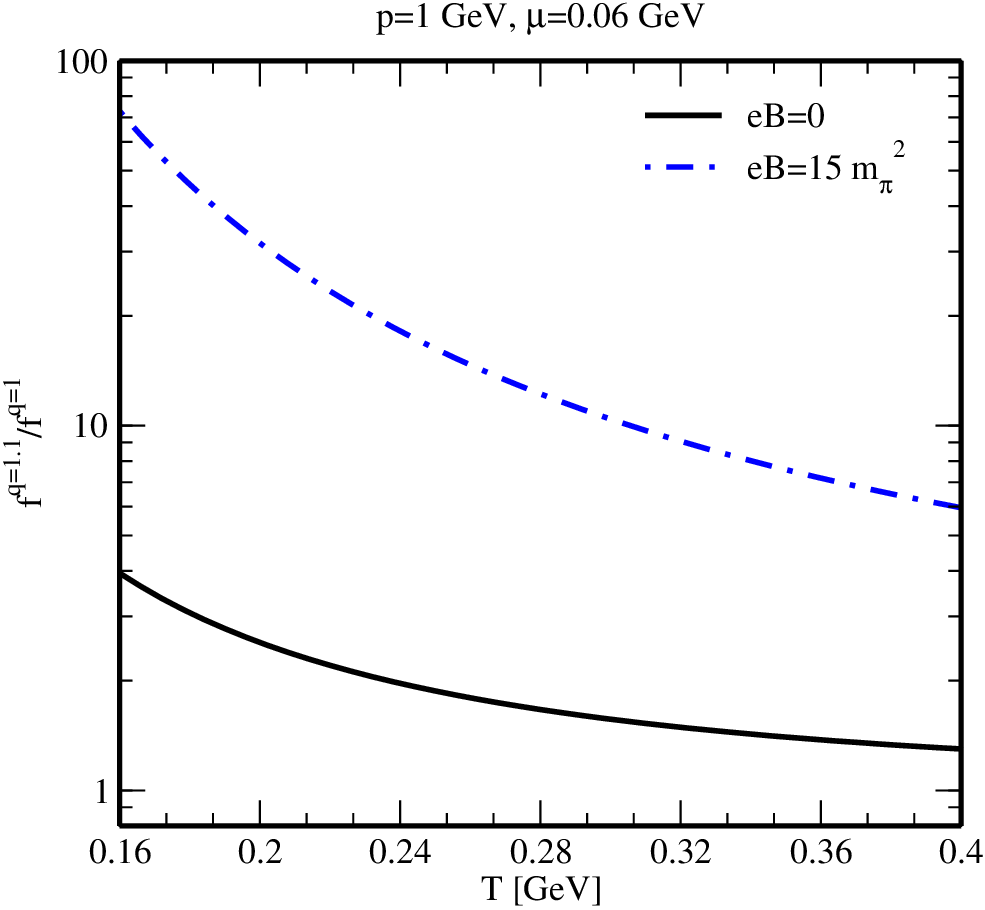}&
\hspace{0.74 cm}
\includegraphics[width=6.3cm]{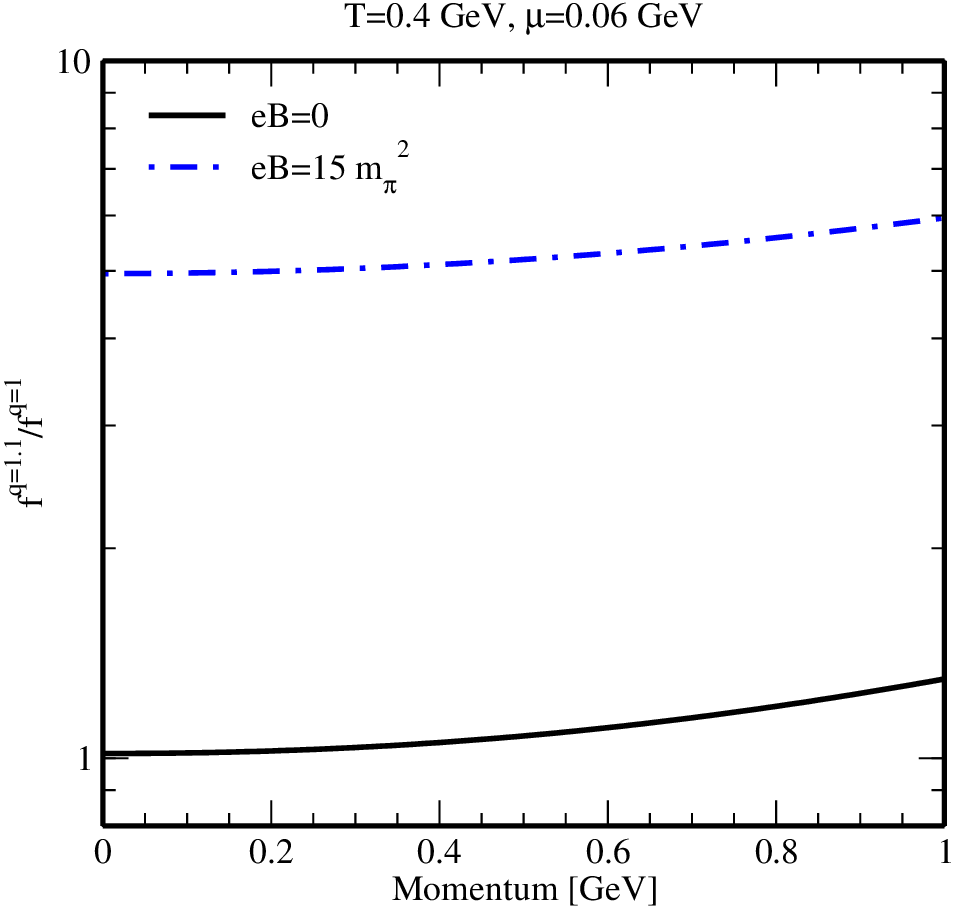} \\
a & b
\end{tabular}
\caption{Variations of the ratio of quark distribution function at $q=1.1$ to that at $q=1$ (a) with temperature and (b) with momentum for zero magnetic field and strong magnetic field cases.}\label{Fig.dfu}
\end{center}
\end{figure}

\begin{figure}[]
\begin{center}
\begin{tabular}{c c}
\includegraphics[width=6.3cm]{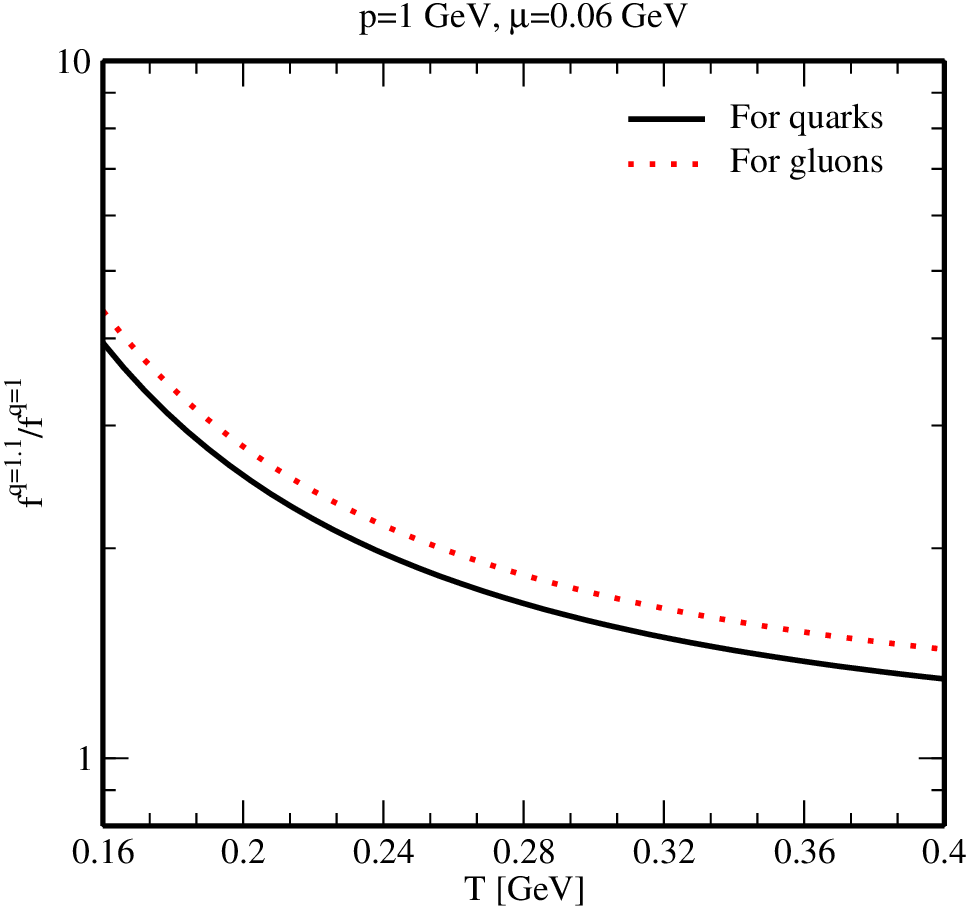}&
\hspace{0.74 cm}
\includegraphics[width=6.3cm]{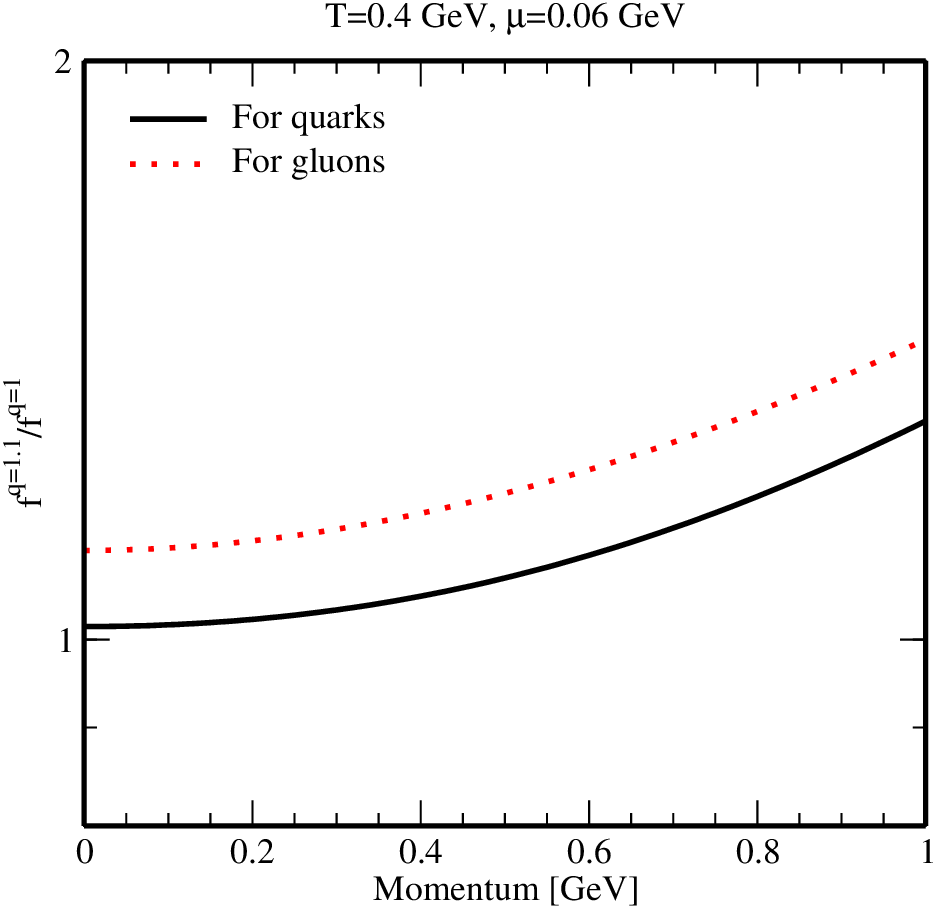} \\
a & b
\end{tabular}
\caption{Variations of the ratio of distribution function at $q=1.1$ to that at $q=1$ (a) with temperature 
and (b) with momentum for quark and gluon cases.}\label{Fig.df}
\end{center}
\end{figure}

As the transport coefficients depend on the features of particle distribution functions, it is relevant to know how the particle distribution functions behave in the pertinent scenario. Figure \ref{Fig.dfu} shows the $u$ quark distribution function at $q=1.1$ scaled with its value at $q=1$ ($f^{q=1.1}/f^{q=1}$) for different conditions of temperature, momentum, chemical potential and magnetic field. This figure deciphers the information on how the nonextensive quark distribution function at $q=1.1$ approaches (deviates) to (from) the corresponding equilibrated distribution function at $q=1$. In particular, figure \ref{Fig.dfu}a depicts the variation of $f^{q=1.1}/f^{q=1}$ with temperature for zero magnetic field and the strong magnetic field cases, whereas figure \ref{Fig.dfu}b shows its variation with momentum. It can be observed that the nonextensive behavior of the quark distribution function is less pronounced at high temperatures and low momenta, because in these cases, the nonextensive Tsallis distribution approaches to the Fermi-Dirac distribution. Figure \ref{Fig.df} compares the nonextensive features of the quark and gluon distribution functions. It is observed that the trend of variation of $f^{q=1.1}/f^{q=1}$ for gluon with temperature and momentum is almost similar to that of the quark. However, the magnitude is larger for the gluon case. It thus indicates that, at a given value of temperature or momentum, the gluons contribute more to the nonextensivity of the medium than the quarks. Throughout the considered temperature and momentum ranges, the nonextensive Tsallis distribution functions for quarks and gluons at $q=1.1$ remain larger than the Fermi-Dirac and Bose-Einstein distribution functions. 

\begin{figure}[]
\begin{center}
\begin{tabular}{c c}
\includegraphics[width=6.3cm]{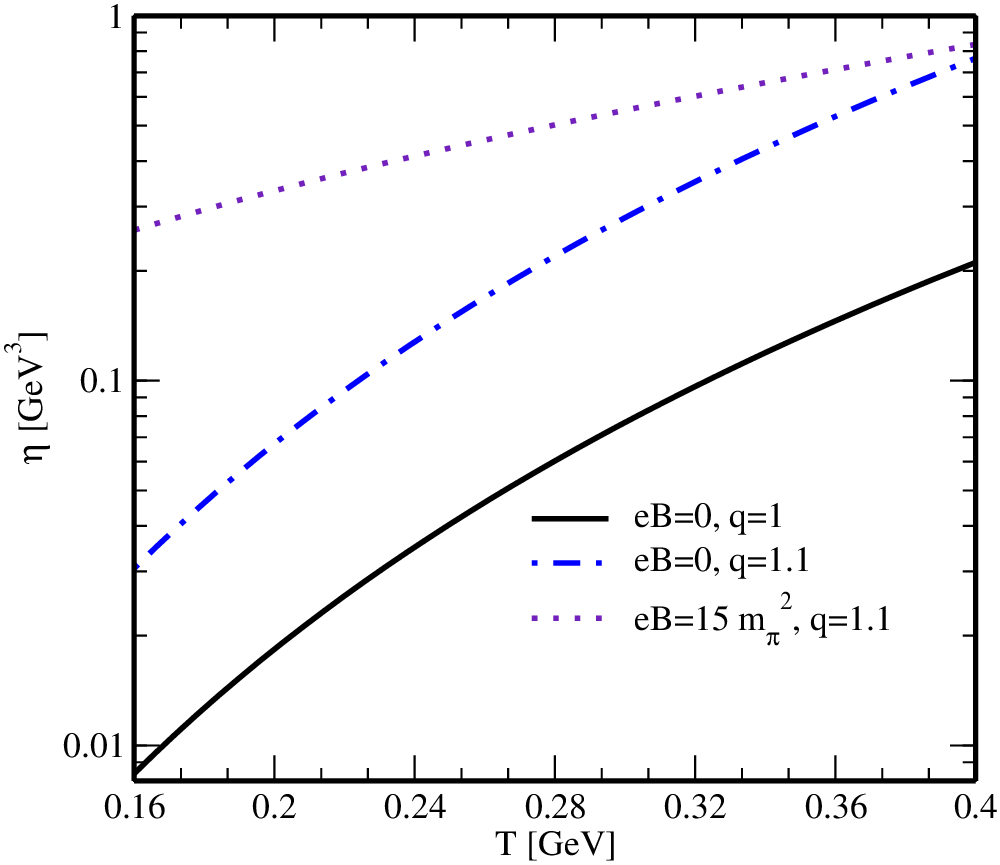}&
\hspace{0.74 cm}
\includegraphics[width=6.3cm]{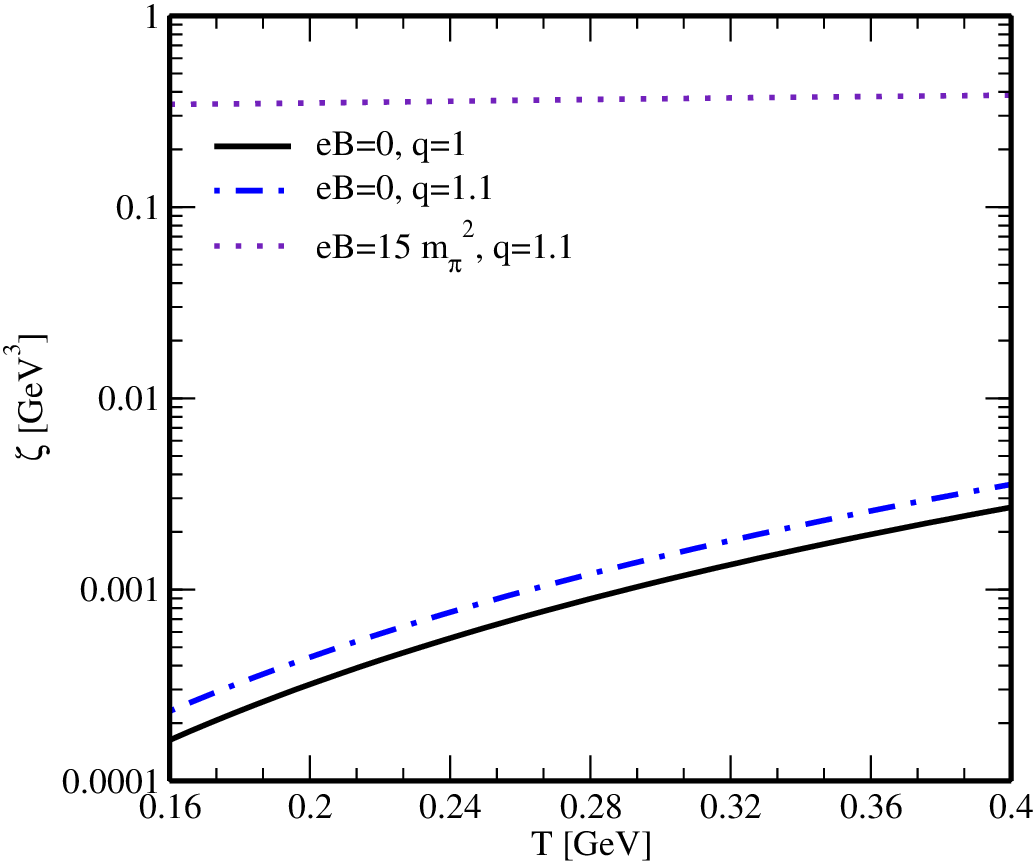} \\
a & b
\end{tabular}
\caption{Variations of (a) the shear viscosity (b) the bulk viscosity with temperature for different 
values of the nonextensive parameter in the presence of a strong magnetic field.}\label{Fig.1}
\end{center}
\end{figure}

Figure \ref{Fig.1} shows the variations of shear ($\eta$) and bulk ($\zeta$) viscosities with temperature. It is observed that the magnitudes of both shear viscosity and bulk viscosity are larger for $q=1.1$ as compared to their counterparts at $q=1$. It explains that the momentum transfer is higher as well as there is an 
enhancement in local pressure fluctuations when the medium is away from equilibrium. When the nonextensive medium comes under the regime of a strong magnetic field, then further increase of both $\eta$ and $\zeta$ can be noticed and this conveys the information that the nonextensive effects are more evident in the presence of a strong magnetic field. This behavior can be mainly attributed to the increase of the nonextensive distribution function with the introduction of a strong magnetic field (figure \ref{Fig.dfu}). Thus the nonextensive Tsallis distribution in the strong magnetic field regime leads to larger values of shear and bulk viscosities than the corresponding Fermi-Dirac and Bose-Einstein thermal distributions. 

Further, it can be seen from figure \ref{Fig.1} that the shear viscosity remains nearly two orders of magnitude larger than the bulk viscosity over the entire range of temperature for both $q=1$ and $q=1.1$ cases. Thus the resistance to any deformation in the system at constant volume is larger than the resistance to change in the volume of the system at constant shape, which is also valid for the nonextensive medium. This indicates that the trend of variation of viscosities with temperature at finite nonextensivity is similar to that for $q=1$. Thus, since the variation of the bulk viscosity with temperature is slower than that of the shear viscosity (especially at high temperatures) for $q=1$, a similar slower variation of the bulk viscosity is also expected for $q=1.1$. As a result, it shows a meagre nonextensive behavior for bulk viscosity. Furthermore, for a nonextensive medium at strong magnetic field, there is much slower variation of the bulk viscosity with temperature as compared to the zero magnetic field case, which is expected because in the strong magnetic field regime, magnetic field acts as the dominant energy scale compared to the temperature, however, as the temperature increases at a fixed magnetic field, the effect of the strong magnetic field limit ($|q_fB| \gg T^2$) gets gradually suppressed and it is more evident for the case of bulk viscosity. 

The extra increase of the viscosities due to the strong magnetic field can be understood as follows: The strong magnetic field restricts the motion of charged particles to one spatial dimension, thus stretching the distribution function along the direction longitudinal to the magnetic field and splitting the phase space integral into longitudinal and transverse parts, where the transverse part contains a factor $|q_fB|$. Further, the quasiparticle model ensures the dependence of the distribution functions on magnetic field, temperature and chemical potential through the modified dispersion relations of partons. Moreover, the partial derivative of pressure with respect to energy density appearing in the bulk viscosity expression incorporates the dependence of magnetic field, contrary to that in the absence of magnetic field. Therefore, $\eta$ and $\zeta$ calculated in a strong magnetic field depend explicitly on magnetic field, chemical potential and temperature, unlike the case in the absence of magnetic field, where they depend on temperature and chemical potential. As a result, in the strong magnetic field limit ($|q_fB| \gg T^2$), where the energy scale associated with the magnetic field is dominant as compared to the energy scales related to the temperature and the chemical potential, $\eta$ and $\zeta$ become more sensitive to the magnetic field. Thus, the emergence of the strong magnetic field enhances the viscosities, in addition to their increase observed due to the nonextensive behavior of the medium. With the increase of temperature, both shear and bulk viscosities increase slowly as compared to the faster increase of their counterparts in the absence of magnetic field. This behavior of $\eta$ and $\zeta$ can be comprehended from the fact that, in the strong magnetic field regime, temperature is a weak energy scale, hence it may leave meagre impact on them, whereas in the absence of magnetic field, temperature is the dominant energy scale, so its effect on the viscosities is more noticeable as compared to that in a strong magnetic field. 

\section{Applications of the viscous properties}
This section is dedicated to the study of some applications of shear and bulk viscosities. In particular, the effects of the nonextensivity on the flow characteristic and on the specific shear and bulk viscosities are studied in subsections 4.1 and 4.2, respectively. 

\subsection{Flow characteristic}
It is possible to comprehend the flow characteristic of the matter by observing the Reynolds number (Re) 
and this is related to the kinematic viscosity (${\eta}/{\rho}$) as
\begin{equation}\label{Rl}
{\rm Re}=\frac{Lv}{\eta/\rho}
~,\end{equation}
where $L$ and $v$ are the characteristic length of the system and the velocity of the flow, respectively. The mass density ($\rho$) can be calculated from the product of the number densities of quarks, antiquarks and gluons with their respective quasiparticle masses as
\begin{equation}\label{M.D.}
\rho=\sum_f m_f\left(n_f+\bar{n}_f\right)+m_gn_g
~.\end{equation}
Thus, the expressions of mass density for a nonextensive hot QCD medium in the absence of magnetic field and in the presence of a strong magnetic field are obtained as
\begin{eqnarray}
\rho_q &=& \frac{1}{2\pi^2}\sum_f m_f g_f \int d{\rm p}~{\rm p}^2\left[f_q+\bar{f}_q\right]+\frac{1}{2\pi^2}m_g g_g\int d{\rm p}~{\rm p}^2f_g, \label{iso.(M.D.)} \\ 
\rho^B_q &=& \frac{1}{4\pi^2}\sum_f m_f g_f|q_fB|\int d p_3\left[f_q^B+\bar{f}_q^B\right]+\frac{1}{2\pi^2}m_g g_g\int d{\rm p}~{\rm p}^2f_g \label{baniso.(M.D.)}
,\end{eqnarray}
respectively. The magnitude of the Reynolds number can convey important information about the fluidity of the system. For ${\rm Re}\gg1$ ({\em i.e.} in the thousands), the kinematic viscosity is much smaller than the product of characteristic length and velocity, thus the nature of the flow is turbulent. For smaller magnitude of the Reynolds number, the medium behaves like a viscous system and the flow remains laminar. It would be interesting to see how the nonextensivity alters the flow characteristic of the medium. For peripheral heavy ion collisions, the velocity of the flow can vary up to the speed of light \cite{Fukutome:PTP119'2008,Csernai:PRC85'2012,McInnes:NPB921'2017}. It is set to be nearly the speed of light, {\em i.e.} $v\simeq c$ for convenience and since in natural units in QCD $c=1$, we have used $v\simeq 1$ in this work. Investigations have shown that the system size depends on the values of the number of participants and the number of collisions in heavy ion collisions for different values of the center-of-mass energy \cite{Zhang:0106242,Miller:ARNPS57'2007} and for the QGP matter, the characteristic size is generally 
chosen to be approximately 3 fm - 4 fm. For the present work, we have set $L=4$ fm. 

\begin{figure}[]
\begin{center}
\includegraphics[width=6.3cm]{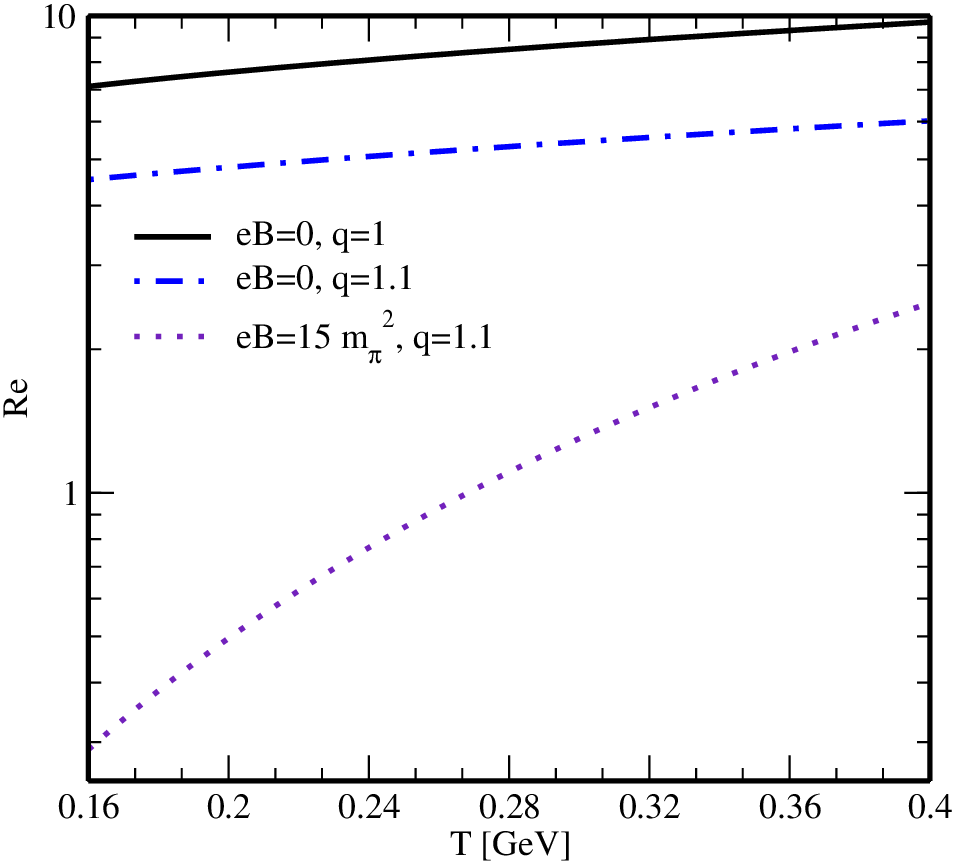}
\caption{Variations of the Reynolds number with temperature for different values 
of the nonextensive parameter in the presence of a strong magnetic field.}\label{Fig.2}
\end{center}
\end{figure}

Figure \ref{Fig.2} depicts the evolution of the Reynolds number with temperature for different values
of $q$ and magnetic field. It can be observed that the Reynolds number increases with the increase of temperature in all cases, however, a decrease in its magnitude is observed when the nonextensive parameter changes from $q=1$ to $q=1.1$. If the medium with $q=1.1$ is exposed to a strong magnetic field, the decrease in the Reynolds number is more evident as it becomes less than unity at low temperatures up to $T\simeq0.26$ GeV. This suggests that the kinematic viscosity dominates over the characteristic length scale of the system 
and the quark-gluon plasma medium becomes more viscous with flow retaining its laminar behavior 
in the said regime. 

\subsection{Specific shear and bulk viscosities}
\begin{figure}[]
\begin{center}
\includegraphics[width=6.3cm]{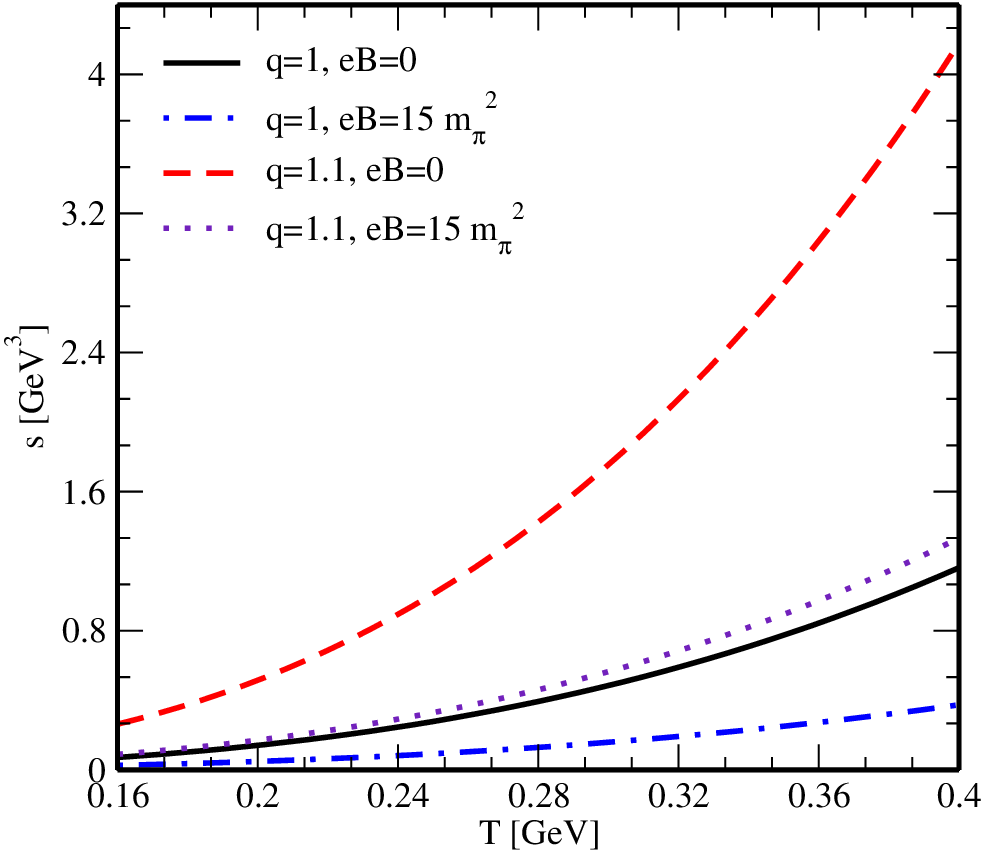}
\caption{Variations of the entropy density with temperature for different values 
of the nonextensive parameter in the absence of magnetic field and in the 
presence of a strong magnetic field.}\label{Fig.q}
\end{center}
\end{figure}

The properties of specific shear ($\eta/s$) and specific bulk ($\zeta/s$) viscosities depend on how the viscosities and the entropy density behave in different conditions of the nonextensivity, magnetic 
field, temperature and chemical potential. The entropy density ($s$) can be evaluated 
from the energy-momentum tensor and baryon density ($n_B$) through the following relation: 
\begin{eqnarray}\label{E.D.}
S=\frac{u_\mu T^{\mu\nu}u_\nu-\sum_{f}\mu_f n_B-\Delta_{\mu\nu}T^{\mu\nu}/3}{T}
~.\end{eqnarray}
Here, the expressions of $n_B$ for a nonextensive hot QCD medium in the absence of magnetic field and in the presence of a strong magnetic field are calculated as
\begin{eqnarray}
(n_B)_q &=& \frac{1}{2\pi^2}\sum_f g_f \int d{\rm p}~{\rm p}^2\left[f_q-\bar{f}_q\right], \\ 
(n_B)^B_q &=& \frac{1}{4\pi^2}\sum_f g_f|q_fB|\int d p_3\left[f_q^B-\bar{f}_q^B\right]
,\end{eqnarray}
respectively. The results of the entropy density at finite nonextensivity ($q=1.1$) and its comparison with $q=1$ case are shown in figure \ref{Fig.q}. It is found that the emergence of the nonextensivity gives an increasing effect to the entropy density at zero magnetic field as well as at strong magnetic field. As the temperature of the medium grows, the deviation of the entropy density at $q=1.1$ from that at $q=1$ increases. However, the deviation is less prominent in the presence of a strong magnetic field, which is due to the 
fact that the phase space gets squeezed to (1+1)-dimensions in a strong magnetic field. This leads to a smaller number of microstates and thus, a reduced entropy density is observed. So, overall, the presence 
of finite nonextensivity makes the medium more disordered, whereas the disorderliness is comparatively 
small at low temperatures. 

\begin{figure}[]
\begin{center}
\begin{tabular}{c c}
\includegraphics[width=6.3cm]{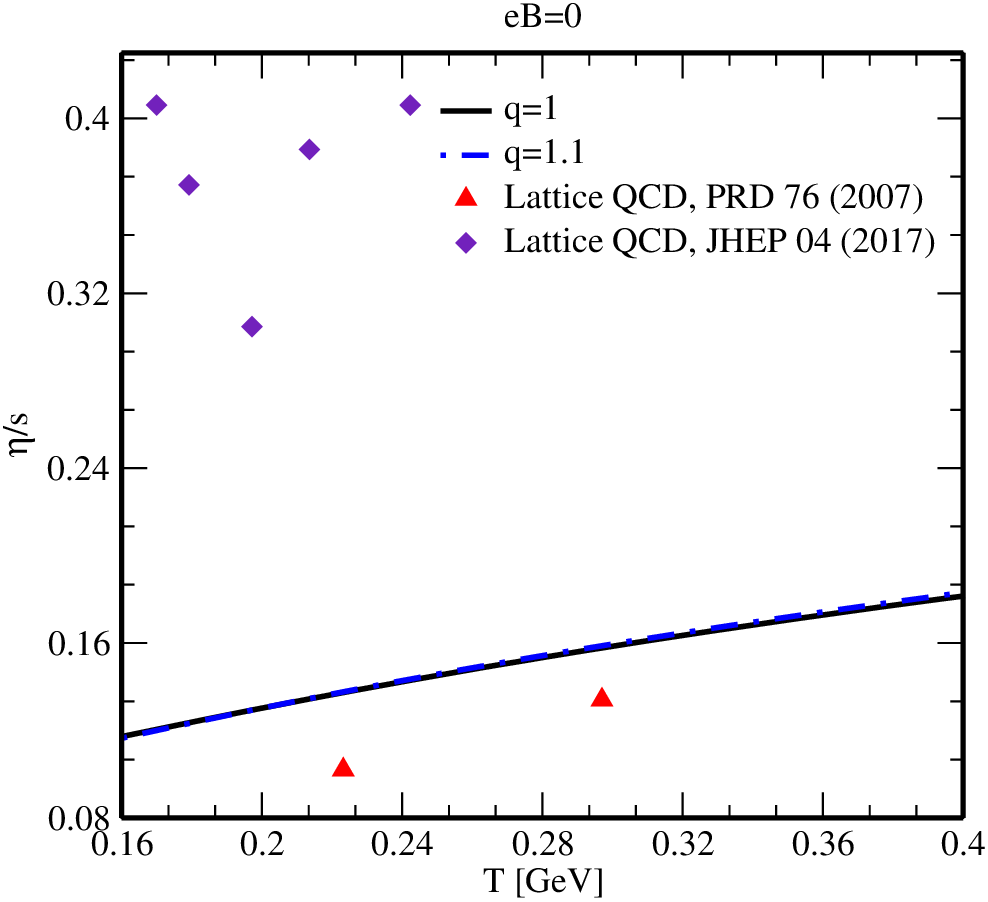}&
\hspace{0.74 cm}
\includegraphics[width=6.3cm]{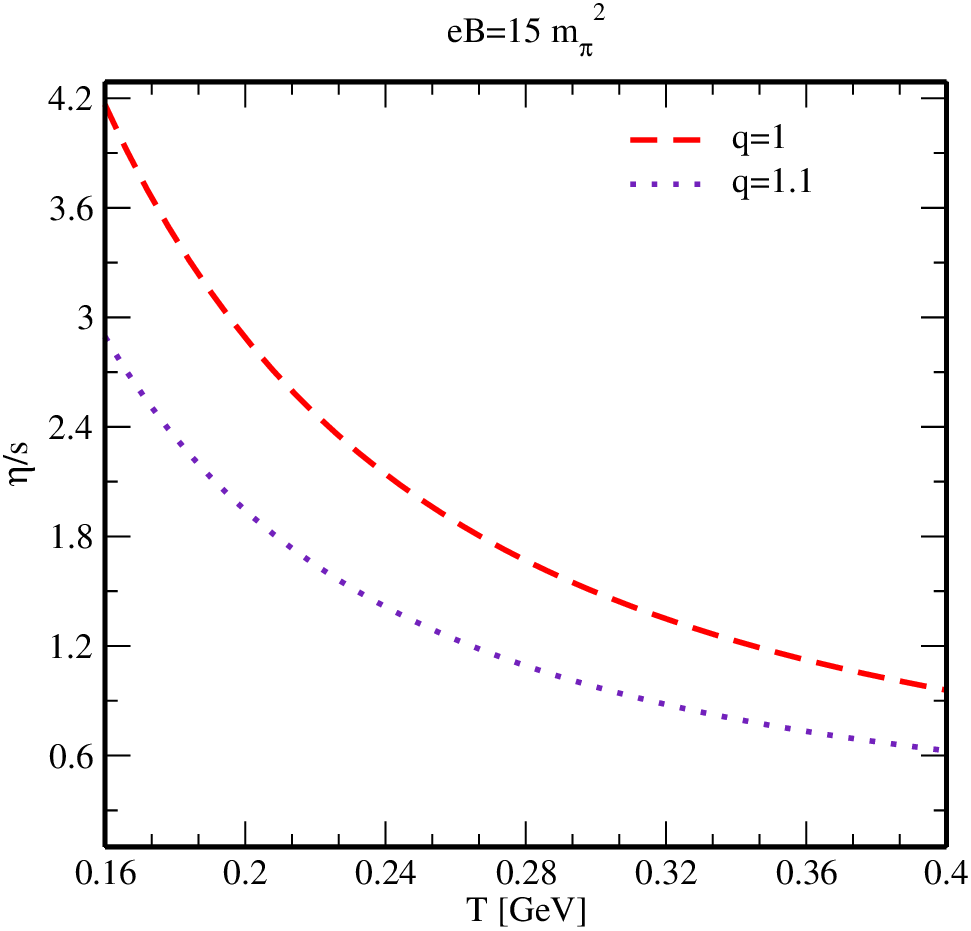} \\
a & b
\end{tabular}
\caption{Variations of the specific shear viscosity with temperature for different values 
of the nonextensive parameter in (a) the absence of magnetic field and (b) the presence 
of a strong magnetic field. Comparison of our result on $\eta/s$ with the lattice QCD results \cite{Meyer:PRD76'2007,Astrakhantsev:JHEP1704'2017} has been made.}\label{Fig.3}
\end{center}
\end{figure}

\begin{figure}[]
\begin{center}
\begin{tabular}{c c}
\includegraphics[width=6.3cm]{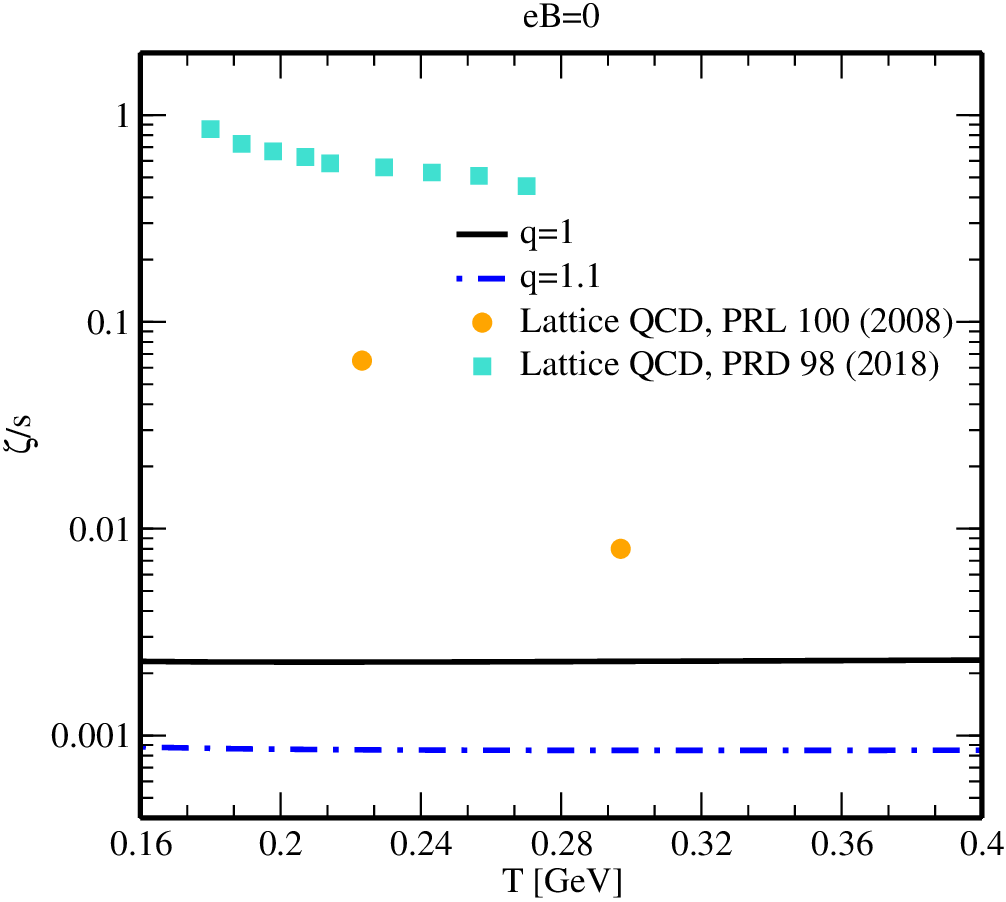}&
\hspace{0.74 cm}
\includegraphics[width=6.3cm]{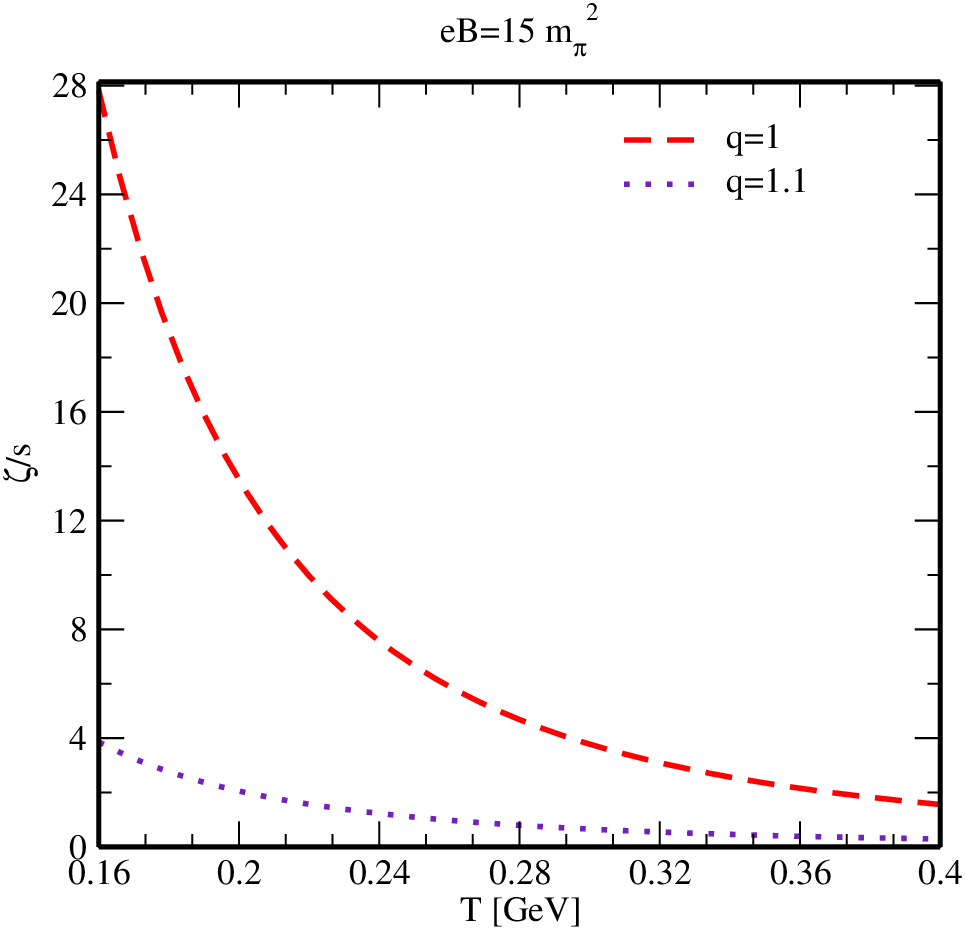} \\
a & b
\end{tabular}
\caption{Variations of the specific bulk viscosity with temperature for different values 
of the nonextensive parameter in (a) the absence of magnetic field and (b) the presence 
of a strong magnetic field. Our result has been compared with the lattice QCD results \cite{Meyer:PRL100'2008,Astrakhantsev:PRD98'2018} on $\zeta/s$.}\label{Fig.4}
\end{center}
\end{figure}

Figures \ref{Fig.3} and \ref{Fig.4} respectively show the variations of $\eta/s$ and $\zeta/s$ 
with temperature for $q=1$ and $q=1.1$. A meagre increase 
of $\eta/s$ (figure \ref{Fig.3}a) and a decrease of $\zeta/s$ (figure \ref{Fig.4}a) due to the nonextensive behavior of the medium are observed. On the other hand, figures \ref{Fig.3}b and \ref{Fig.4}b 
respectively illustrate a noticeable decrease of both $\eta/s$ and $\zeta/s$ due to the influence 
of the nonextensivity in the strong magnetic field regime, with the decrease of latter one being larger in magnitude than the former one. This decrease of $\eta/s$ at finite nonextensivity can be understood from the fact that the increase of $\eta$ is smaller in magnitude than the increase of $s$. Compared to $\eta/s$, the decrease of $\zeta/s$ at finite nonextensivity is comparatively larger, because the nonextensive behavior of the medium gives an increasing effect to the bulk viscosity, which is smaller in magnitude than that of the shear viscosity, thus an overall larger decreasing impact on $\zeta/s$ is observed. It is observed that, at zero magnetic field, the ratio $\eta/s$ is nearer to the conjectured lower bound $1/(4\pi)$, and the ratio $\zeta/s$ is closer to the conformal limit due to the nonextensive behavior of the medium. 

We have also compared our results with lattice QCD calculations. According to the lattice calculation for the SU(3) pure gauge model \cite{Nakamura:PRL94'2005}, the upper bound for $\eta/s$ of QGP was estimated to be 1, and our result on $\eta/s$ at finite nonextensivity lies slightly below this bound for the temperature range 0.16 GeV - 0.4 GeV. However, its magnitude in the presence of a strong magnetic field exceeds the lattice result. According to the lattice work in ref. \cite{Meyer:PRD76'2007}, $\eta/s$ approaches 0.102 at $T=1.24T_c$ and 0.134 at $T=1.65T_c$, whereas our nonextensive results of $\eta/s$ lie slightly above these values at the said temperatures. A very small magnitude of $\zeta/s$ ($<$0.15) except near $T_c$ had been observed in the lattice calculation in ref. \cite{Meyer:PRL100'2008}, which became extremely small with the increasing temperature above $T_c$ and our nonextensive result is observed to be smaller than this lattice result on $\zeta/s$. In ref. \cite{Astrakhantsev:JHEP1704'2017}, the SU(3)-gluodynamics had been applied on the lattice through the Backus-Gilbert method to study the shear viscosity and it reported the range of $\eta/s$ to be approximately 0.40 - 0.30 for the temperature range $0.17 ~ {\rm GeV} - 0.20 ~ {\rm GeV}$, whereas our nonextensive result on $\eta/s$ ranges 0.12 - 0.13 for the said temperature range, thus, even with the nonextensive effect, our result on $\eta/s$ remains below that of the lattice result. The temperature dependence of bulk viscosity using the SU(3)-gluodynamics on lattice had been explored in ref. \cite{Astrakhantsev:PRD98'2018} and it reported very small values of $\zeta/s$ for $T\geq1.1T_c$, which lie above our results for $q=1$ and $q=1.1$ in the same temperature range. Generally, $\zeta/s$ vanishes for a conformal QCD medium, thus the decrease of $\zeta/s$ at finite nonextensivity takes the medium a bit closer to the conformal symmetry. 

\section{Conclusions}
In this work, we studied how the nonextensivity of the medium affects its viscous properties in the absence as well as in the presence of a strong magnetic field. The shear ($\eta$) and bulk ($\zeta$) viscosities of hot QCD matter were calculated in the kinetic theory approach by solving the relativistic Boltzmann transport equation within the nonextensive Tsallis mechanism. The interactions among particles were manifested by their quasiparticle masses. The effects of the nonextensivity on some applications of the aforesaid viscosities, such as the flow characteristic, specific shear and specific bulk viscosities were also explored. We observed that the introduction of the nonextensivity enhances the values of $\eta$ and $\zeta$, and these enhancements are comparatively larger in the presence of a strong magnetic field. Thus, it confirmed that the nonextensive behavior of the medium amplifies the momentum transfer as well as the fluctuations in local pressure. Throughout the considered temperature range, $\eta$ remains larger than $\zeta$, thus indicating larger momentum transfer across the layer than along the layer. The viscous nature of the hot QCD matter becomes more evident for $q=1.1$ due to the decrease of the Reynolds number in the similar environment. This observation also affirmed the laminar nature of the flow. It was found that the ratio $\eta/s$ is nearer to the conjectured lower bound at finite nonextensivity. Further, from the observation on the ratio $\zeta/s$, it was found that, due to the emergence of the nonextensivity, the medium approaches more towards the conformal symmetry. 

\section{Acknowledgments}
One of the present authors (S. R.) is thankful to the Indian Institute of Technology Bombay 
for the Institute postdoctoral fellowship and S. D. acknowledges the SERB Power 
Fellowship, SPF/2022/000014 for the support on this work.

\end{document}